\documentclass{article}
\usepackage{amsmath, amssymb, graphics, setspace}
\usepackage{graphicx,graphics}
\usepackage{latexsym,epsf,rotate}
\usepackage{hyperref}
\usepackage{xcolor}
\usepackage{epsfig, epstopdf}

\def\beq{\begin{equation}} 
\def\eeq{\end{equation}}
\def\bea{\begin{eqnarray}}
\def\eea{\end{eqnarray}}
\def\eqref#1{eq.~(\ref{eq:#1})}

\def\vslash#1{\mbox{/\llap #1}} 
\def\nn{\nonumber}   

\begin{document}

\begin{titlepage}

\begin{flushright}
  { \bf IFJPAN-IV-2023-3  \\
    November 2023
  \hskip 3 mm
}
\end{flushright}

\vspace{0.1cm}
\begin{center}
  {\Large \bf Electron-positron, parton-parton and
    photon-photon production of $\tau$-lepton pairs: 
    anomalous magnetic and electric dipole moments spin effects}   
\end{center}

 \vspace{0.1cm} 
\begin{center}
  {\bf  Sw.~Banerjee$^a$,  A.Yu.~Korchin$^{b,c,d}$, E.~Richter-Was$^e$ and Z.~Was$^{d}$}\\
{\em  $^a$University of Louisville, Louisville, Kentucky 40292, USA} \\
{\em  $^b$NSC Kharkiv Institute of Physics and Technology, 61108 Kharkiv, Ukraine} \\
{\em  $^c$V.N.~Karazin Kharkiv National University, 61022 Kharkiv, Ukraine} \\
{\em $^d$Institute of Nuclear Physics Polish Academy of Sciences, PL-31342 Krakow, Poland}\\
{\em  $^e$Institute of Physics, Jagiellonian University, ul. Lojasiewicza 11, 30-348 Krakow, Poland}\\ 
\end{center}
\vspace{.1 cm}

\begin{center}
{\bf   ABSTRACT  }   
\end{center}

Anomalous contributions to the electric and magnetic dipole moments of the $\tau$ lepton
from new physics scenarios have brought renewed interest in the development of new charge-parity violating signatures
in $\tau$-pair production at Belle II energies, and also at higher energies of the Large Hadron Collider and the Future Circular Collider.
In this paper, we discuss the effects of spin correlations, including transverse degrees of freedom,  
in the $\tau$-pair production and decay.
These studies include calculating analytical formulas, obtaining numerical results,
and building semi-realistic observables sensitive to the transverse spin correlations
induced by the dipole moments of the $\tau$ lepton.
The effects of such anomalous contributions to the dipole moments are introduced on top of precision simulations
of $e^-e^+ \to \tau^-\tau^+$, $q\bar{q} \to \tau^-\tau^+$ and $\gamma\gamma \to \tau^-\tau^+$ processes,
involving multi-body final states. The  $\tau$ decays are simulated along with radiative 
corrections, in particular electroweak box contributions of  $WW$ and $ZZ$ exchanges are taken into account. 
Respective extensions of the Standard Model amplitudes and the reweighting algorithms are implemented
into the {\tt KKMC} Monte Carlo, which is used to simulate $\tau$-pair production in $e^-e^+$ collisions,
and the {\tt TauSpinner} program, which is used to reweight events with $\tau$ pair produced in $pp$ collisions.

\vfill %
\vspace{0.1 cm}

\vspace*{1mm}
\bigskip

\end{titlepage}

\section{Introduction}

Electric and magnetic dipole moments of the $\tau$ lepton are sensitive to the violation of fundamental symmetries,
such as the charge-parity (CP) violation~\cite{Ramsey:1982pq, Cheng:1983mh, Bernreuther:1989kc}.
Recent measurements of dipole moments of the $\tau$ lepton at the Belle experiment~\cite{Belle:2021ybo},
as well as observation of $\gamma\gamma \to \tau^-\tau^+$ production at the hadron colliders~\cite{ATLAS:2022ryk,CMS:2022arf}
have brought renewed interest in phenomena of electric and magnetic dipole moments of the $\tau$ lepton.
Deviation of measured values of the magnetic moment of the muon~\cite{Muong-2:2021ojo} from predictions of the Standard Model (SM), 
and possibly enhanced contributions from New Physics (NP) models to the magnetic moment of the $\tau$ lepton, makes these studies important and of contemporary interest.
These NP contributions to magnetic moments are expected to be proportional to the square of the mass of the corresponding lepton.
Several NP scenarios introduce dark weakly-interacting scalars or vector states
accompanying production of heavy fermions e.g. $\tau$ leptons.
Such states and other new virtual particles via loop corrections,
can be sources of anomalous contributions to the electric or magnetic dipole moments of the $\tau$ lepton,
as mentioned in Ref.~\cite{Bernreuther:1996dr, Eidelman:2007sb}  and references therein. 

The {\tt KKMC} Monte Carlo {(MC)} has been used to generate events of the SM $e^-e^+ \to \tau^-\tau^+$ process
for several decades, and is now enriched with several additional processes to describe
NP effects potentially observable at $e^-e^+$ collider and the Belle II experiment~\cite{Banerjee:2021hke}.
In particular, recent development in Ref.~\cite{Banerjee:2022sgf} allows one
to include anomalous contributions to electric and magnetic dipole moments of the $\tau$ lepton,
in case of low energy $\tau$-pair production. In {\tt KKMC}, all spin correlations are rigorously implemented, including quantum entanglement effects.
This is the case for configurations where an arbitrary number of hard bremsstrahlung photons can be present.
It can also provide valuable benchmarks for spin effects in $e^-e^+ \to \tau^- \tau^+$ process, with $\tau$ decays included. 

The {\tt TauSpinner} program is a convenient tool to study {and prepare} observables sensitive to the NP effects at
hadron colliders.  {Its purpose is to introduce, with the help of weights, small effects on top of event
  samples of the SM content. It is imperative, as in case of {\tt KKMC}, to evaluate
  if the precision of the SM simulation is sufficient for the user's needs.
Then, {\tt TauSpinner} may be used to introduce NP and/or spin effects.
The samples may be generated with the help of general purpose MC generators
like {\tt Pythia} \cite{Bierlich:2022pfr}, {\tt Sherpa} \cite{Sherpa:2019gpd} or with the help of other, sophisticated generator setup.}  
{In its earlier versions, {\tt TauSpinner} was used to introduce the longitudinal spin effects,
 in the case of the Drell-Yan processes or the Higgs production with decays into $\tau$ leptons,
 at the Large Hadron Collider (LHC)~\cite{Czyczula:2012ny}.}  Later, they were extended to applications
of the NP interactions \cite{Kalinowski:2016qcd} in the hard processes,
in which lepton pairs are accompanied with one or two hard jets \cite{Kalinowski:2016qcd},
and to include the transverse spin effects~\cite{Przedzinski:2014pla}.
The list of applications was also extended to incorporate the electroweak {(EW)} 
effects~\cite{Richter-Was:2018lld,Richter-Was:2020jlt}.

Let us now introduce some terminology, which is used throughout the paper.
In the reaction $\gamma \gamma \to \tau^- \tau^+$, where the $\tau$ lepton couples with on-shell photons,
we include the anomalous magnetic and electric dipole moments of the $\tau$ lepton. 
In the processes $e^- e^+ \to \tau^- \tau^+$ and $q \bar{q} \to \tau^- \tau^+$,
where the couplings of the $\tau$ lepton with the photon or the $Z$-boson are virtual,
one usually introduces form-factors.
For the photon exchange, we have form-factors for the anomalous magnetic dipole moment ($a$) and the electric dipole moment ($d$),
while for the $Z$-boson exchange, we have form-factors for the anomalous weak magnetic dipole moment ($a_w$) and the weak electric dipole moment ($d_w$). 
In case of on-shell photons with virtuality of $q^2\sim 0$, the electromagnetic form-factors reduce to the corresponding dipole moments. 
Similarly, for on-shell  $Z$-boson with virtuality $q^2\sim M_Z^2$, the weak form-factors reduce to the corresponding weak dipole moments (see, e.g. Ref.~\cite{Hollik:1998vz}).    

All these form-factors are connected to the chirality flipping operators. 
The terms proportional to electric and weak electric form-factors are also $CP$ violating. 
Anomalous magnetic form-factor also receives a contribution from radiative corrections in the SM,
which we separate from the NP term. As the electric dipole form-factor in the SM is highly suppressed,
one can assume that this form-factor comes exclusively from NP contributions.       
 
The current experimental results and constraints for these dipole moments (or form-factors in some cases),
together with theoretical predictions in the SM, are presented in Table~\ref{tab:tau_dipoles}.

\begin{table}[tbh]
\begin{center}
\begin{tabular}{c  c  c  }       
\hline
   & SM prediction & Experiment~\cite{ParticleDataGroup:2022pth}  \\
\hline
$a$ &  $1.17721(5) \times 10^{-3}$~\cite{Eidelman:2007sb} &
$ -0.052 <  a < 0.013 $~\cite{DELPHI:2003nah}   \\ 
$a_w$	& $ - (2.10 + i \, 0.61) \times 10^{-6} $~\cite{Bernabeu:1994wh}
& ${\rm Re}(a_w) < 1.14  \times 10^{-3}$~\cite{ALEPH:2002kbp}  \\
&  &  ${\rm Im} (a_w) < 2.65  \times 10^{-3}$ at 95\% C.L.~\cite{ALEPH:2002kbp} \\ 													
$d$   &   $ - 7.32 \times 10^{-38}$ e$\cdot$cm~\cite{Yamaguchi:2020dsy}
&   $(-1.85 < {\rm Re}(d) < 0.61) \times 10^{-17}$ e$\cdot$cm~\cite{Belle:2021ybo}\\
& & $(-1.03 < {\rm Im}(d) < 0.23)  \times 10^{-17}$ e$\cdot$cm~\cite{Belle:2021ybo}\\
 $d_w$    &  --  & ${\rm Re}(d_w) < 5.01  \times 10^{-18}$  e$\cdot$cm~\cite{ALEPH:2002kbp}\\ 
     &    &     ${\rm Im}(d_w) < 11.15  \times 10^{-18}$ e$\cdot$cm~\cite{ALEPH:2002kbp}\\
\hline
\end{tabular}
\caption{Current status of predictions and
  measurements of anomalous magnetic ($a$), 
weak anomalous magnetic ($a_w$), electric ($d$) 
and weak electric ($d_w$) dipole moments/form-factors, of the $\tau$ lepton,
quoted at 95\% confidence level. } 
\label{tab:tau_dipoles}
\end{center}
\end{table}
					
In this paper, we focus on the effects of dipole moments/form-factors on spin correlations
in the $\tau$-pair production and decay. Quantifying those effects requires explicit calculation of the matrix elements
including longitudinal and transverse spin components and their correlations. 

First, we consider the SM case, because it determines the nature of interfering contributions for analysis of 
anomalous dipole moments due to NP. 
{The SM amplitude is evaluated in the improved Born approximation (IBA).
The theoretical basis of IBA is formulated in Ref.~\cite{Bardin:1999yd}, and our approach is explained in 
Ref.~\cite{Richter-Was:2018lld}.
It allows us to separate the EW correction into pure  QED part and the remaining one, sometimes called the genuine weak part.
However, we shorten the name to simply refer to this as the EW correction.
Numerically these corrections are included in IBA and rely on the EW {\tt Dizet} library~\cite{Bardin:1999yd}, which is installed  in {\tt KKMC}.
The EW corrections are introduced with form-factor corrections to the SM couplings and propagators, which enter the IBA amplitude  used for calculation of EW weights.
They represent complete ${\cal O}(\alpha)$ EW corrections with QED contributions removed but augmented with carefully selected dominant higher-order terms.
The $WW$ and $ZZ$ box-diagram contributions, which are numerically important above the $WW$ and $ZZ$ threshold, are thus accounted for.
This approach was very successful in analyses of LEP I precision physics, and later extended for use  in 
Tevatron and LHC precision physics.}
  
In the second step, we consider the anomalous dipole moments themselves.
These studies include calculation of the analytical formulas,
validations of the tools, and an attempt to build 
semi-realistic observables sensitive to  transverse spin correlations induced
by the dipole moments. Impact on the size of cross sections for $\tau$-pair production is addressed marginally.
{Note, that effect of dipole moments in the radiative decay $Z \to \tau^- \tau^+ \gamma$ based on LEP data was discussed in Ref.~\cite{Grifols:1990ha}, and for $\gamma \gamma \to \tau^- \tau^+$ 
in peripheral Pb+Pb collisions in Ref.~\cite{Dyndal:2020yen}}. 

Technically, the effect of anomalous dipole moments  can be introduced on top of precision simulations
of $e^-e^+ \to \tau^-\tau^+$, $q\bar{q} \to \tau^-\tau^+$ and $\gamma\gamma \to \tau^-\tau^+$ processes,
involving multi-body final states and correlations between $\tau$ decay products.
The MC techniques using event-by-event weights, are convenient for this purpose.
The {\tt KKMC} MC is used for simulating spin-correlation effects in $\tau$ pair produced in $e^-e^+$ collisions,
while the {\tt TauSpinner} program is used for reweighting events where $\tau$ pairs are produced in $pp$ collisions.

Our paper is organized in the following manner.
In Sec.~\ref{sec:KKMC-th}, we present analytic calculation of the spin correlation matrix,
which include effects of the $\tau$-lepton anomalous magnetic and electric dipole form-factors. 
Analytic formulas, described in Sec.~\ref{sec:transverse}, are useful for implementation of transverse spin correlations
for each of the parton level processes $q \bar{q} \to \tau^- \tau^+$,   $e^- e^+ \to \tau^- \tau^+$  and $\gamma \gamma \to \tau^- \tau^+$.
In Sec.~\ref{sec:reweighting}, we recall main points of the {\tt KKMC} and {\tt TauSpinner} reweighting algorithms for inclusion of the dipole moments.
Discussions on the possible test observables and some numerical results are collected in 
Sec.~\ref{sec:numerical}. Sec.~\ref{sec:summary} closes the paper with a summary and outlook.

\section{Amplitudes and spin correlations}
\label{sec:KKMC-th}

In Ref.~\cite{Banerjee:2022sgf}, formulas for including anomalous magnetic and electric dipole form-factors 
in elementary $2 \to 2$ parton processes of $\tau$-pair production were discussed.
The elementary parton-level and improved Born-level formulas were presented there for
$e^-e^+ \to \tau^-\tau^+$ process, including details of reference
frames\footnote{A comment on the direction of reference frames is in place here.  Formally speaking, the interchange of reaction name between  $e^-e^+ \to \tau^-\tau^+$
  and  $e^+e^- \to \tau^+\tau^-$ is purely conventional and not of  physics content. The choice however does imply
  which among the incoming $e^-$ or $e^+$ particle is taken as the first beam (analogously if $\tau^-$ or $\tau^+$ 
	is taken as the first outgoing lepton) in the positive $\hat{z}$ direction.
  The second choice  was used in some of older projects. Nowadays, the choice of $e^-$ 
  is made more frequently, as the more energetic beam in asymmetric colliders. Due to overlapping conventions, a sign mismatch may occur.
  In our programs, a rotation by an angle of $\pi$ around the $\hat{y}$ axis may be necessary
  before the anomalous contributions are inserted into the other parts of the code. 
  }
used in spin-correlation matrix calculation. Those formulas  are however limited to  low energies, where contribution
of $Z$-boson exchange can be neglected in specific applications.

For higher energy $e^-e^+ \to \tau^-\tau^+$, and
also for $q \bar{q}  \to \tau^-\tau^+$ parton level processes at the LHC,
inclusion of the contribution due to the exchange of $Z$-boson is necessary. 
For very high $\tau$-pair virtuality of more than $~160$ GeV, the contributions from doubly 
resonant $WW$ and $ZZ$ boxes need to be taken into account as well, following for example,  
Refs.~\cite{Richter-Was:2018lld, Arbuzov:2020coe} 
on {\tt TauSpinner} and {\tt KKMC}, where EW loop corrections installation is documented. 
Here we extend results of Ref.~\cite{Banerjee:2022sgf}, and derive the formulas including $Z$ 
and $\gamma$ exchanges and their interference.

To complete the picture we also derive the formulas for  $\gamma \gamma  \to \tau^-\tau^+$
processes in the so-called light-by-light scattering approximation, which recently became of high interest in the analysis of heavy-ion beams collisions at the LHC~\cite{ATLAS:2022ryk, CMS:2022arf}.

Let us start by introducing formalism for $f_i \,  \bar{f_i} \to \tau^-\tau^+$ process, where
$i=$ $e$, $u$, $d$ or $\gamma$. For the $\gamma \gamma$ initial state, the symbol $\bar{f_i}$ 
refers to the second incoming photon. Thus, in our generalized notation, we consider:

\begin{equation}
f_i (k_1) + \bar{f_i} (k_2) \to \tau^- (p_-) + \tau^+ (p_+),
\label{eq:001}
\end{equation}
where the four-momenta satisfy the conservation relation: $k_1 +k_2 = p_- + p_+ $. 

In the center-of-mass (c.m.) frame, the components of the momenta are  
\begin{eqnarray}
&& p_- =(E, \vec{p}), \qquad \;  p_+ =(E, \, -\vec{p}), \qquad  \vec{p} = (0, \, 0, \, p), 
\nonumber \\ 
&& k_1 = (E, \, \vec{k}) , \qquad k_2  = (E, \, -\vec{k}),  \qquad 
\vec{k} =  (E \, \sin \theta, \, 0, \, E \, \cos \theta ), 
\label{eq:002}
\end{eqnarray}
so that the $\hat{z}$ axis is along the momentum $\vec{p}$, the reaction plane $\hat{x} \hat{z}$ is defined 
by the momenta $\vec{p}$ and $\vec{k}$, and the $\hat{y}$ axis is along $\vec{p} \times \vec{k}$. 
Here, $E=\sqrt{s}/2$,  
$p = \beta E$ is the $\tau$-lepton three-momentum, $\beta = (1-4 m_\tau^2/ s)^{1/2}$ is the velocity,
$m_\tau$ is the mass of the $\tau$ lepton and $s = (p_- + p_+)^2$. 
{The mass of initial lepton or quark is neglected hereafter.}

The quantization frames of $\tau^-$ and  $\tau^+$ are connected to this reaction frame by 
the appropriate boosts along the $\hat{z}$ direction.  Note that the $\hat{z}$ axis  is parallel to 
momentum of $\tau^-$ but anti-parallel to momentum of $\tau^+$. The beams momenta  reside 
in the  $\hat{x} \hat{z}$ plane. Only the reaction frame, the $\tau^-$ and the $\tau^+$ rest 
frames are used for calculations throughout the paper. The vector indices used in the formulas: 1, 2, 3, 4 
correspond to $\hat{x},~ \hat{y},~ \hat{z},~\hat{t} $ directions, respectively.

\subsection{The  $f_i \bar f_i \to \tau^- \tau^+$,  $i=e, u, d $ case}
\label{subsec:e-e_q-qbar}

\vspace{0.2cm}
\centerline{{\it Dipole form-factors and improved Born approximation}}
\vspace{0.2cm}

We consider the $\gamma \tau \tau$ electromagnetic vertex to have the following structure:
\begin{equation}
\Gamma^\mu_\gamma (q)  = -ie {{Q_\tau}}  \Big\{ \gamma^\mu + \frac{\sigma^{\mu \nu} q_\nu}{2 m_\tau} \,  
\big[ i  A(s)   + B(s)  \gamma_5  \big] \Big\},
\label{eq:003}
\end{equation}
where  $\sigma^{\mu \nu}= \frac{i}{2} [\gamma^\mu, \, \gamma^\nu]$,  $e$ is the charge of the positron,
{{ $Q_\tau$ is the charge of $\tau$ lepton in units of $e$},
$A(s) =F_2(q^2)$ is the Pauli form-factor, and $B(s) = F_3(q^2)$ is the electric dipole form-factor, which depend 
on $s = q^2$, where $q=k_1+k_2=p_- + p_+$.  At the on-shell photon point, $A(0)$ is an 
anomalous magnetic dipole moment $a$, while $B(0)$ is related to the $CP$ violating electric 
dipole moment $d$,  
\begin{equation}
A(0) = a = \frac{1}{2} (g-2),    \qquad \quad B(0)  = \frac{2m_\tau}{e {Q_\tau}} d,
\label{eq:004}
\end{equation}
where $g$ is the gyromagnetic factor. 

To separate SM contribution from NP we explicitly include   
the QED correction to $A(s)$ in the first order in $\alpha = e^2/(4 \pi)$~\cite{Berestetskii:1982qgu}
\begin{equation}
A(s)_{QED}=\frac{\alpha m_\tau^2}{\pi  s  \beta} \, \Big(  \log \frac{1-\beta}{1+\beta} + i \, \pi \Big).
\label{eq:016}
\end{equation}
Therefore we have  
\begin{equation}
A(s) = A(s)_{QED} + A(s)_{NP}, \quad \quad B(s) = B(s)_{NP}
\label{eq:017}
\end{equation}  
neglecting very small contribution to $B(s)$ in the SM. 

The  $Z \tau \tau$ vertex is chosen to have the following structure:
\begin{equation}
\Gamma^\mu_Z (q)  = {{ -i}} \frac{g_Z}{2} \, \Big\{ \gamma^\mu (v_{\tau} - \gamma_5 a_{\tau}) 
+ \frac{\sigma^{\mu \nu} q_\nu}{2 m_\tau} \,  
\big[ i  X(s)  + Y(s)  \gamma_5  \big] \Big\},
\label{eq:018}
\end{equation} 
where $g_Z =e/(s_W c_W)$  = $2 M_Z (\sqrt{2} G_F)^{1/2} $,  $s_W \equiv \sin \theta_W$, 
$c_W \equiv \cos \theta_W$;  $\theta_W$ is the weak mixing angle and 
$G_F = 1.1663788(6) \times 10^{-5}$ GeV$^{-2}$~\cite{ParticleDataGroup:2022pth} is the Fermi constant.  
The vector and axial-vector couplings 
for the $\tau$ lepton are $v_{\tau} = -1/2 + 2 s_W^2 $ and $a_{\tau} =-1/2$, respectively,
$X(s)$ is the weak anomalous magnetic form-factor, and $Y(s)$ is related to the $CP$ violating weak
electric form-factor\footnote{Sometimes these form-factors are defined with the additional
factor $(2 c_W s_W)^{-1}$~\cite{Bernabeu:1993er,Bernabeu:1994wh,Hollik:1998vz}.}.

The $X (M_Z^2)$  was evaluated for the SM in Ref.~\cite{Bernabeu:1994wh}. 
Using this calculation, in our notation  
 $X(M_Z^2)_{SM} = - (2.10 + i \, 0.61) \times 10^{-6} \times (2 s_W c_W) $.
It is rather small, and this contribution is not explicitly included in our code,
as elaborated in Sec.~\ref{sec:reweighting}, but can easily be included if necessary.
In this description, only the anomalous component is included, but $X(M_Z^2)_{SM} $ can be introduced into the code,  as additional part of $X (M_Z^2)$.
   
For the initial state fermion $f_i$, the electric charge in unit of $e$
is denoted as $Q_i$, while the vector and axial-vector constants 
are denoted   by $v_i$ and $a_i$, respectively,
as shown in Table~\ref{tab:Z-couplings}.

\begin{table}[tbh]
\begin{center}
\begin{tabular}{c  c  c c  }       
\hline
 fermion  & $Q_i$ & $v_i$  & $a_i$  \\
\hline
 $e^- \, (\mu^-, \tau^-)$ & -1 &  $-\frac{1}{2} +2 s_W^2$  & $-\frac{1}{2}$  \\ 
	                        &      & $ -0.03783 \pm0.00041$                  &   $-0.50123 \pm 0.00026$    \\ 
\hline													
 $u$  & $+\frac{2}{3}$ & $ +\frac{1}{2} - \frac{4}{3} s_W^2 $ & $+\frac{1}{2}$ \\
         &   & $+0.266 \pm 0.034$  & $+0.519^{+0.028}_{-0.038}$ \\
\hline				
 $d $  & $-\frac{1}{3}$ & $-\frac{1}{2} +\frac{2}{3} s_W^2 $ & $ -\frac{1}{2}$  \\
          &   &$-0.38^{+0.04}_{-0.05} $  & $-0.527^{+0.040}_{-0.028}$ \\   
\hline
\end{tabular}
\caption{Electric charge (in units of $e$),  vector and axial-vector couplings for  leptons, $u$ and $d$ quarks.
  The measured values of the effective couplings~\cite{ParticleDataGroup:2022pth} are also 
shown under the values predicted by the SM. It is worthwhile to note the change of sign between the leptons and the $u$ quark.} 
\label{tab:Z-couplings}
\end{center}
\end{table}

To simplify calculations, it is convenient to use the Gordon identities for the matrix elements 
of $\Gamma_\gamma^\mu (q)$ and $\Gamma_Z^\mu (q)$. This allows one to rewrite the respective currents as
\begin{eqnarray}
 \bar{u} (p_-) \,  \Gamma^\mu_\gamma (q)  \, v(p_+) &=& -ie {{Q_\tau}} \, \bar{u} (p_-) 
\Big\{ \gamma^\mu (1+A(s))  
\nonumber \\
&& + \frac{(p_+ - p_-)^\mu}{2m_\tau} \big[A(s) - i B(s) \gamma_5 \big] \Big\}  v(p_+),     \label{eq:020} \\
\bar{u}(p_-) \, \Gamma^\mu_Z (q) \, v(p_+) &=& {{-i}} \frac{g_Z}{2} \,  \bar{u}(p_-) 
\Big\{ \gamma^\mu ((v_{\tau}+X(s))  - \gamma_5 a_{\tau})  \nonumber \\ 
&& +  \frac{(p_+ - p_-)^\mu}{2m_\tau} \big[ X(s) - i Y(s) \gamma_5 \big] \Big\}  v(p_+).
\label{eq:021}
\end{eqnarray}

{{ Next we introduce radiative corrections following Refs.~\cite{Bardin:1999yd,Richter-Was:2018lld}. 
The amplitude, called ${\cal M}^{IBA}$, for arbitrary initial ($i$) and final ($f$) fermions can be written as 
\begin{eqnarray}
\label{eq:IBA}
{\cal M}^{IBA} &=& \frac{e^2 Q_f Q_i}{s} V_{fi} (s,t) \, \gamma_\mu \otimes \gamma^\mu  \\
&& +
\Bigl(\frac{g_Z}{2} \Bigr)^2 \frac{Z_{fi}(s,t)}{d(s)} \, \gamma_\mu 
[v_i (s,t) - a_i \gamma_5] \otimes  \gamma^\mu [v_f (s,t) - a_i \gamma_5], \nonumber
\end{eqnarray} 
where the following shortcut notation is used:  operator on the left of $\otimes$ is sandwiched between 
the spinors  $ \bar{v}(k_2) \ldots u(k_1)$, and operator on the right of $\otimes$ is sandwiched 
between the spinors $\bar{u}(p_-) \ldots v(p_+)$.  In Eq.~(\ref{eq:IBA}) $Q_i$ and $Q_f$ 
are the charges of initial and final fermions  in units of $e$.} }

{{Furthermore,  $t = (k_1 - p_-)^2$,  $d(s) = s - M_Z^2 + i s \, \Gamma_Z / M_Z$ 
with running $Z$-boson decay 
width and  $Z_{f i} (s, t)$ is the normalization correction for the $Z$-boson propagator.
$v_{i}(s,t)$ and $v_f (s,t)$ are modified vector couplings 
\begin{equation}
v_i (s,t) = T_{3 i} - 2 Q_i s_W^2 K_i (s,t), \qquad v_f (s,t) = T_{3  f} - 2 Q_f s_W^2 K_f (s,t),
\label{eq:vivf}
\end{equation}
where $T_{3}$ is the 3rd component of the fermion weak isospin.}  
{{In Eq.~(\ref{eq:IBA}), the factor $V_{fi} (s,t) $ includes the terms:
\begin{equation}
V_{fi} (s,t)  = \Gamma_{vp} (s) + \Bigl(\frac{g_Z}{e}\Bigr)^2 s_W^4 Z_{fi} (s, t)  \frac{s}{d(s)}
[K_{f i}(s,t) - K_f (s,t) K_i (s,t)], 
\label{eq:effective_gamma}
\end{equation} 
where $\Gamma_{vp} (s)$ includes re-summed vacuum-polarization loop contributions  
to the photon propagator,  and the complex EW form-factors $K_i(s,t)$,  $K_f (s,t)$ and $K_{fi}(s,t)$ are defined in Ref.~\cite{Richter-Was:2018lld}.}

{{Eq.~(\ref{eq:IBA}) represents improved photon exchange amplitude with running QED constant 
and improved $Z$-boson exchange which include:
\begin{itemize}
\item[--] corrections to the photon propagator coming from the vacuum-polarization loops,
\item[--] corrections to the $Z$-boson propagator and couplings embedded in the form-factors $Z_{fi}(s,t)$,
$K_i(s,t), \, K_f (s,t)$ and $K_{fi}(s,t)$, 
\item[--] contributions from the $WW$ and $ZZ$ box diagrams also included in the form-factors,
\item[--] mixed ${\cal O}(\alpha \alpha_s, \, \alpha \alpha_s^2, \ldots)$ corrections originating from gluon 
insertions in the self-energy loop diagrams.
\end{itemize}} 
 
{{The EW form-factor corrections are available in {\tt Dizet} library \cite{Bardin:1999yd}.
For further details we refer to Ref.~\cite{Richter-Was:2018lld}.
Note that corrections in IBA become numerically sizable at high-energies, well above $Z$-boson peak.  
At the lower energies one can use the amplitude
in which all the form-factors and $\Gamma_{vp} (s)$ are replaced by 1.} 

{{ Next we add the amplitude due to the dipole form-factors of the $\tau$ lepton, introduced 
in (\ref{eq:003}) and (\ref{eq:018}).
Using identities (\ref{eq:020}) and (\ref{eq:021}), one obtains  
\begin{eqnarray}
\label{eq:DM}
&& {\cal M}^{DM} = \frac{e^2 Q_f Q_i}{s} V_{f i} (s,t) \, \gamma_\mu \otimes [A \gamma^\mu  +
\frac{(p_+ - p_-)^\mu}{2m_\tau} (A - i B \gamma_5)]  \\
&& +
\Bigl(\frac{g_Z}{2} \Bigr)^2 \frac{Z_{f i}(s,t)}{d(s)} \, \gamma_\mu 
[v_i (s,t) - a_i \gamma_5] \otimes  [X \gamma^\mu  +\frac{(p_+ - p_-)^\mu}{2m_\tau} 
(X - i Y \gamma_5)], 
\nonumber
\end{eqnarray}
and the total amplitude is $ {\cal M} = {\cal M}^{IBA} + {\cal M}^{DM}$.}

\vspace{0.2cm}
\centerline{{\it Spin-correlation matrix}}
\vspace{0.2cm}

We consider production of the polarized $\tau^-$ and $\tau^+$ leptons, which are characterized   
by the following polarization 3-vectors in their rest-frames, respectively:
\begin{equation}
\vec{s}^{\, -} = (s^-_1, \, s^-_2, \, s^-_3),  \qquad  \vec{s}^{\,+}=   (s^+_1, \, s^+_2, \, s^+_3),
\label{eq:008}
\end{equation}
where the Cartesian components are defined with respect to the chosen frame.  
For convenience, we also use unity as the 4-th components of the spin vectors: 
\begin{equation}
s^- = (s^-_1, \, s^-_2, \, s^-_3, \, 1 ), \qquad s^+ =   (s^+_1, \, s^+_2, \, s^+_3, \, 1 ).
\label{eq:009}
\end{equation}

The square of the matrix element averaged over all the polarization states of the initial fermions  can be written as:
\begin{equation}
|{\cal M}|^2 = |{\cal M}_\gamma|^2 + |{\cal M}_Z|^2 + 2 {\rm Re} ({\cal M}_\gamma^* \, {\cal M}_Z) ,
\label{eq:022}
\end{equation}  
which determines the differential cross section taking into account the spin degrees of freedom of the $\tau$ leptons as:
\begin{equation}
\frac{d \sigma }{d \Omega} (f \, \bar f \to \tau^- \tau^+)= \frac{\beta}{64 \pi^2   s } 
|{\cal M}|^2  = \frac{\beta}{64 \pi^2   s }  \sum_{i, j=1}^4 \,  R_{i, j} 
\, s^-_i  s^+_j ,
\label{eq:023}
\end{equation}
where $R_{i, j} $ is explained below.
   
We rewrite Eq.~(\ref{eq:023}) in the following more convenient form:
\begin{eqnarray}
&& \frac{d \sigma }{d \Omega} (f \, \bar f \to \tau^- \tau^+) =
\frac{d \sigma }{d \Omega} (f \, \bar f \to \tau^- \tau^+)\Big|_{{\rm unpol}} \nonumber \\
&& \times \frac{1}{4}  \Big( 1  + 
\sum_{i=1}^3 r_{i,4} \,  s^-_i  + \sum_{j=1}^3  r_{4, j} \,  s^+_j  +  
\sum_{i, j = 1}^3 r_{i, j} \, s^-_i  s^+_j  \Big) .
\label{eq:023a}
\end{eqnarray} 
The  elements of spin-correlation matrix $r_{i, j} \equiv R_{i, j}/R_{44}$ and  the $\tau^-, \tau^+$ 
polarization states are respectively $r_{i, 4} \equiv R_{i,4}/R_{44}$, $r_{4, j} \equiv R_{4,j}/R_{44}$, where $i, j \le 3$. 

The cross section for unpolarized $\tau$'s is expressed through $R_{44}$: 
\begin{equation}
\frac{d \sigma }{d \Omega} (f \, \bar f \to \tau^- \tau^+) \Big|_{{\rm unpol}}
= \frac{\beta}{16 \pi^2   s } \, R_{44}. 
\label{eq:023b} 
\end{equation}
 
If $\tau$ decays are taken into account,  
the vectors defining $\tau^-$ and $\tau^+$ density matrices, 
$s^-_i$ and $s^+_j$ in Eq.~(\ref{eq:023a}), are replaced  respectively by the polarimetric vectors,
depending on the $\tau$-decay matrix elements, $h^-_i$ and $h^+_j$.

Furthermore, each contribution in $R_{i, j}$ can be broken down as: 
\begin{equation}
R_{i, j} =    R^{(\gamma)}_{i, j} + R^{(Z)}_{i, j} + R^{(\gamma Z)}_{i, j},
\label{eq:024}
\end{equation}
where $R^{(\gamma)}_{i, j} $, $R^{(Z)}_{i, j}$ and  $R^{(\gamma Z)}_{i, j}$ represent contributions  
due to $\gamma$ exchange, $Z$-boson exchange and $\gamma \, Z$ interference, respectively. 
In the following, we include only terms linear in {the dipole} form-factors.   
The quantization of $\tau^\mp $ are performed in their respective rest frames.
These frames differ from the reaction frame, defined by Eq.~(\ref{eq:002}), by the boosts along $\tau^\mp$ directions. 

{For energies below and around $Z$-boson peak, we show  results 
  with radiative corrections switched off, as they can be there safely neglected (incorporated into 
	redefinition  of constants).
  For higher energies, the phenomenological picture would be far more complicated
  also because of experimental detection/reconstruction criteria, and possibly
  initial state bremsstrahlung photons of $p_T \sim m_\tau$ getting lost in the beam-pipe. That is why,
  this regime remain out of scope of the present work}\footnote{{Note that in contrary to
    Belle II energies or c.m. energy corresponding to the Z peak, at higher energies
    presence of initial state photons of $p_T$ comparable or higher than the tau mass and at the
    same time lost in the beam-pipe are not disfavored. This complicates construction of useful
    for our purposes observables. At the Z-peak energies, such photons are disfavored because
    then amplitudes are not resonant. At low energies, the ratio of tau mass to beam energy is much
    larger than the detector minimal acceptance angle.
  }}. 

Let us start by recalling results of Ref.~\cite{Banerjee:2022sgf} for $ R^{(\gamma)}_{i,j}$ in case of 
$e^- e^+ \to \tau^- \tau^+$ process, extended now to also include quarks in the initial state:   
\begin{eqnarray}
&& R^{(\gamma)}_{11} = \frac{e^4 \, Q_i^2}{4  \gamma ^2 \, N_i} 
\big[4  \gamma ^2 \, {\rm Re}(A(s)) +\gamma ^2+1 \big] \sin^2(\theta), 
\label{eq:A1} \\
&& R^{(\gamma)}_{12} = -R^{(\gamma)}_{21} =   
\frac{e^4 \, Q_i^2 }{2 \, N_i} \beta \sin^2(\theta)  \, {\rm Re}(B(s)),   \nonumber \\
&& R^{(\gamma)}_{13} = R^{(\gamma)}_{31} =  
\frac{e^4 \, Q_i^2}{4 \gamma \, N_i } \big[ ( \gamma ^2+1 ) {\rm Re}(A(s))+1 \big] 
\sin(2 \theta),
\nonumber \\ 
&& R^{(\gamma)}_{22} = -\frac{e^4 \, Q_i^2 }{4 \, N_i}  \beta^2 \sin^2(\theta), \nonumber \\
&& R^{(\gamma)}_{23} = -R^{(\gamma)}_{32} = 
- \frac{e^4 \, Q_i^2 }{4 \, N_i} \beta \, \gamma  \sin(2 \theta) \, 
{\rm Re}(B(s)),  \nonumber \\
&& R^{(\gamma)}_{33} = \frac{e^4 \, Q_i^2 }{4 \gamma^2 N_i} \big\{ \big[ 4 \gamma ^2 \, 
{\rm Re}(A(s)) +  \gamma^2 + 1 \big] 
\cos^2(\theta) + \beta^2 \gamma^2 \big\}, \nonumber \\
&& R^{(\gamma)}_{14} = -R^{(\gamma)}_{41} 
=  \frac{e^4 \, Q_i^2}{4 \, N_i} \beta \, \gamma \sin(2 \theta) \, 
{\rm Im}(B(s)),  \nonumber \\
&& R^{(\gamma)}_{24} = R^{(\gamma)}_{42} =  
\frac{e^4 \, Q_i^2 }{4 \, N_i} \beta^2 \, \gamma   \sin(2 \theta) \, {\rm Im}(A(s)),  \nonumber \\
&& R^{(\gamma)}_{34} = - R^{(\gamma)}_{43} = 
- \frac{e^4 \, Q_i^2 }{2 \, N_i} \beta  \sin^2(\theta)  \, {\rm Im}(B(s)),   \nonumber \\
&& R^{(\gamma)}_{44} = \frac{e^4 \, Q_i^2  }{4 \gamma ^2 \, N_i} 
\big[4  \gamma^2 \, {\rm Re}(A(s))  + \beta^2 \gamma^2  \cos^2(\theta) + \gamma^2 +1 \big]. \nonumber
\end{eqnarray}
The color factor $N_i$ in Eqs.~(\ref{eq:A1}) is equal to $N_c=3$ and $N_i=1$
for the $q \bar{q}$ and $e^- e^+$ initial state, respectively,
while $\gamma = \sqrt{s}/(2 m_\tau)$ denotes the Lorentz factor. 

Next, we extend the results of Ref.~\cite{Banerjee:2022sgf} taking into account the $Z$-boson exchange.
Let us start with the component $R^{(Z)}_{i, j}$ that takes into account only the contribution due to the $Z$-boson exchange,
and is the most significant contribution in the vicinity of the $Z$-peak. This contribution is calculated in terms of the $\tau$-lepton 
Lorentz factor $\gamma$, the velocity $\beta$, and the angle $\theta$ between the momenta of incoming $f_i$ and $\tau^-$ lepton.  
We express the square of the modulus of the $Z$-boson propagator as: 
\begin{equation}
D_Z(s) = (s - M_Z^2)^2 + M_Z^2 \Gamma_Z^2, 
\label{eq:025}
\end{equation}
where $M_Z$ and $\Gamma_Z$ are the mass and decay width of the $Z$-boson, respectively, and obtain:

\begin{eqnarray}
&& R_{11}^{(Z)}  = \frac{g_Z^4 m_\tau^4 \gamma ^2}{4  N_i  D_Z (s)}  
\sin^2 (\theta) \big\{ \left(1- \gamma ^2\right) a_{\tau}^2+ v_{\tau} \bigl[ 4 \gamma ^2 \text{Re}(X(s)) \nonumber \\
&& +\left(1+\gamma
^2\right) v_{\tau} \bigr] \big\}    \left(a_i^2+v_i^2\right), \label{eq:026}  \\ 
&&
R_{12}^{(Z)}=  - \frac{g_Z^4 m_\tau^4 \beta  \gamma ^4}{2  N_i  D_Z(s)}    { \sin^2 (\theta) 
\big[ \text{Im}(X(s)) a_{\tau} -\text{Re}(Y(s)) v_{\tau} \big] \left(a_i^2+v_i^2\right) }, \nonumber \\  
&&
R_{22}^{(Z)}= \frac{g_Z^4 m_\tau^4 \gamma ^2}{4  N_i  D_Z(s)}  \left(\gamma ^2 -1 \right) 
{\sin}^2 (\theta) \left(a_{\tau}^2 - v_{\tau}^2\right) \left(a_i^2+v_i^2\right), \nonumber \\
 && 
R_{13}^{(Z)}=  \frac{g_Z^4 m_\tau^4 \gamma ^3}{4  N_i  D_Z(s)}
 \big\{  4 \,  \beta \,  \text{sin}(\theta) \, a_i \left[ \gamma ^2 \text{Re}(X(s)) a_{\tau} +
\left(-\gamma ^2 \text{Im}(Y(s))+a_{\tau} \right) v_{\tau} \right]  v_i \nonumber \\
&&+\text{sin}(2 \theta) \left[ \left(1-\gamma ^2\right) \text{Im}(Y(s))    
a_{\tau} +v_{\tau} \left(\left(1+\gamma ^2\right) \text{Re}(X(s))+v_{\tau} \right)\right]
\left(a_i^2+v_i^2\right) \big\}, \nonumber \\ 
&& 
R_{23}^{(Z)} =  -   \frac{ g_Z^4 m_\tau^4 \gamma ^3}{2  N_i  D_Z(s)} 
\text{sin}(\theta) \big\{  \beta  \gamma ^2 \text{Re}(Y(s)) \left[ v_{\tau} \text{cos}(\theta) 
(a_i^2 +v_i^2) +2 \beta  a_{\tau} a_i v_i \right] \nonumber \\
&& +\text{Im}(X(s)) \left[ 2 \left(-1+\gamma ^2\right) a_i v_{\tau} v_i 
 + \beta  \gamma ^2 \text{cos}(\theta) a_{\tau} \left(a_i^2+v_i^2\right)\right] \big\} ,  \nonumber \\
&&
R_{33}^{(Z)}  = \frac{g_Z^4 m_\tau^4 \gamma ^2}{8  N_i  D_Z(s)}
 \big\{ 16 \beta  \gamma ^2 \text{cos}(\theta) a_{\tau} \, a_i \, v_{\tau} \, v_i +\left(\gamma ^2 -1 \right) \nonumber \\  
&& \times (3+\text{cos}(2 \theta) ) \,
a_{\tau}^2 \, \left(a_i^2+v_i^2\right)+\left(-1+3 \gamma ^2 +\left(1 +\gamma ^2 \right) 
\text{cos}(2 \theta)\right) v_{\tau}^2 \left(a_i^2+v_i^2 \right) \nonumber \\ 
&& +8 \gamma
^2 \text{cos}(\theta) \text{Re}(X(s)) \left[ v_{\tau} \text{cos}(\theta) (a_i^2+v_i^2) +2 \beta  a_{\tau} \, a_i \, v_i
\right] \big\}, \nonumber \\
&&
R_{14}^{(Z)}  = 
\frac{g_Z^4 m_\tau^4 \gamma ^3}{2  N_i  D_Z(s)}
 \text{sin}(\theta) \big\{ \text{Im}(Y(s)) \bigl[ \beta \,  \gamma^2 \, v_{\tau} \, \text{cos}(\theta) (a_i^2 + v_i^2)
\nonumber \\ 
&&  +2 \left(\gamma ^2 -1\right) a_{\tau} a_i v_i  \bigr] 
-v_{\tau} \left(2 a_i \, v_{\tau} \, v_i +\beta  \text{cos}(\theta) a_{\tau} \left(a_i^2+v_i^2\right) \right)  \nonumber \\ 
&& -\text{Re}(X(s))
\left[ 2 \left(1+\gamma ^2 \right) a_i \, v_{\tau} \, v_i +\beta  \gamma ^2 \text{cos}(\theta) a_{\tau} 
\left(a_i^2+v_i^2\right) \right] \big\},  \nonumber \\ 
&&
R_{24}^{(Z)} =
\frac{ g_Z^4 m_\tau^4 \gamma ^3}{2  N_i  D_Z(s)}
 \text{sin}(\theta) \big\{ \text{Im}(X(s)) \bigl[ \left(\gamma ^2 -1 \right) v_{\tau} \, \text{cos}(\theta) (a_i^2 + v_i^2) 
\nonumber \\
&& +2 \beta  \gamma^2 a_{\tau} \, a_i \, v_i \bigr]
+\text{Re}(Y(s)) \left[ 2 \beta  \gamma ^2 a_i \, v_{\tau} \, v_i +\left(\gamma ^2 -1 \right)
\text{cos}(\theta) a_{\tau} \left(a_i^2+v_i^2 \right) \right] \big\}, \nonumber \\ 
&&
R_{34}^{(Z)} = 
\frac{g_Z^4 m_\tau^4 \gamma ^2}{4  N_i  D_Z(s)}
 \big\{-4  a_i \,  v_i \, \text{cos}(\theta) (\left(\gamma ^2 -1 \right)  a_{\tau}^2 +  \gamma ^2  v_{\tau}^2) 
\nonumber \\ 
&&-\beta \gamma ^2 (3+\text{cos}(2 \theta) ) \, a_{\tau} \, v_{\tau} \left(a_i^2+v_i^2 \right)
-\gamma ^2 \bigl[ 2 \beta  \,  \text{Im}(Y(s)) \text{sin}^2(\theta) v_{\tau} \left(a_i^2+v_i^2\right)
\nonumber \\
&&+\text{Re}(X(s)) \left(  8 \text{cos}(\theta) a_i \,  v_{\tau} \, v_i +\beta  (3+\text{cos}(2 \theta)) a_{\tau} 
\left(a_i^2+v_i^2\right) \right) \bigr] \big\}, \nonumber \\
&&
R_{44}^{(Z)} =
\frac{g_Z^4 m_\tau^4 \gamma ^2}{8  N_i  D_Z(s)}
 \big\{ 16 \beta  \gamma ^2 \text{cos}(\theta) \, a_{\tau} \, a_i \, v_{\tau} \, v_i      
 	+\left(\gamma ^2 -1\right) (3 +\text{cos}(2 \theta) )    \nonumber \\
&& \times a_{\tau}^2 \left(a_i^2+v_i^2\right)+\bigl(1+3 \gamma ^2  
 +\left(\gamma ^2 -1\right) \text{cos}(2 \theta) 
\bigr) v_{\tau}^2 \left(a_i^2+v_i^2\right) \nonumber \\ 
&&
+8 \gamma^2 \, \text{Re}(X(s)) \bigl[a_i^2 v_{\tau} +2 \beta \,  \text{cos}(\theta) a_{\tau} \, a_i \, v_i +v_{\tau} \, v_i^2 
\bigr] \big\}, 
\nonumber
\end{eqnarray}

and the remaining components can be obtained using:
\begin{equation}
R_{j,i}^{(Z)} = R_{i,j}^{(Z)}  \, \Big|_{X(s) \to X(s), \, Y(s) \to -Y(s)} \quad \text{for} \; \;  i \ne j.
\label{eq:027}
\end{equation}

If only the $e^-e^+$ initial state is considered, Eqs.~(\ref{eq:026}) can be simplified.
It simplifies further if one takes into account 
that the vector coupling for the leptons $v_{\tau} \approx -0.038 $ is numerically quite small. Here, we will keep only the terms
linear in $v_{\tau}$,  but neglect the terms $v_{\tau} \, X(M_Z^2)$ and $v_{\tau} \, Y(M_Z^2)$.   
Using these approximations, we obtain for the $Z$-boson contribution (for brevity we denote below $X \equiv X(M_Z^2)$ and $Y \equiv Y(M_Z^2)$) at the energy corresponding to the $Z$-boson peak: 
\begin{eqnarray}
&& 
R^{(Z)}_{11}= -R^{(Z)}_{22}=   - \frac{g_Z^4 \, a_{\tau}^4 \,  \beta^2 \, M_Z^2}{64 \, \Gamma _Z^2} \, 
\text{sin}^2 (\theta),  \label{eq:R_Z} \\
&& 
R^{(Z)}_{12}= R^{(Z)}_{21}= - \frac{g_Z^4 \, a_{\tau}^3 \, \beta \, M_Z^2  }{32 \, \Gamma_Z^2} \,
 \text{sin}^2(\theta) \, \text{Im}(X),  \nonumber \\
&&
R^{(Z)}_{13}=- R^{(Z)}_{31}=  -  \frac{g_Z^4 \, a_{\tau}^3 \, \beta^2 \, M_Z^2}{64 \, \Gamma _Z^2} \, 
\gamma \, \text{sin}(2 \theta) \, \text{Im}(Y) ,   \nonumber \\
&&
R^{(Z)}_{23}= R^{(Z)}_{32}= - \frac{g_Z^4 \, a_{\tau}^3 \, \beta \,  M_Z^2}{64 \, \Gamma_Z^2} \,
 \gamma \, \text{sin}(2 \theta) \,  \text{Im}(X) ,  \nonumber \\
&&
R^{(Z)}_{14}=R^{(Z)}_{41}= -\frac{g_Z^4 \, a_{\tau}^3 \,  \beta \,  M_Z^2 }{64 \, \Gamma_Z^2} \,
\gamma \, \text{sin}(2 \theta)  \, \bigl[ \text{Re}(X)  + {v_{\tau}}\gamma^{-2} \bigr] , \nonumber \\
&&
R^{(Z)}_{24}= - R^{(Z)}_{42}=  \frac{g_Z^4 \, a_{\tau}^3 \, \beta^2 \,  M_Z^2}{64 \, \Gamma _Z^2} \,
\gamma \, \text{sin}(2 \theta) \, \text{Re}(Y) ,   \nonumber \\
&&
R^{(Z)}_{34}= R^{(Z)}_{43} =
 - \frac{g_Z^4 \, a_{\tau}^3 \, \beta \, M_Z^2 }{32 \, \Gamma_Z^2} \,
 \big\{ \bigl(1 +\text{cos}^2(\theta) \bigr) \, \bigl[ v_{\tau} + \text{Re}(X) \bigr]  \nonumber \\ 
&& \qquad \qquad \qquad \quad + 2 v_{\tau} \, \beta \, \cos(\theta) \big\}, 
 \nonumber \\
&&
R^{(Z)}_{44}=R^{(Z)}_{33}= \frac{g_Z^4 \, a_{\tau}^4 \, \beta^2 \, M_Z^2 }{64 \, \Gamma_Z^2} \, 
\, \bigl( 1+\text{cos}^2(\theta) \bigr) , \nonumber
\end{eqnarray}   
where  $\gamma=M_Z /(2 m_{\tau}) \approx 25.7$ and $\beta \approx 1$. 

Finally, let us present the contribution from $\gamma \, Z$ interference. Close to the $Z$-peak, and using the same approximations as above,
we obtain for the case of incoming $e^-e^+$ (for brevity we denote below $A \equiv A(M_Z^2)$ and $B \equiv B(M_Z^2)$):

\begin{eqnarray}
&& R^{(\gamma Z)}_{11} = R^{(\gamma Z)}_{22}= 0,  
\label{eq:R_gamma-Z}\\
&& R^{(\gamma Z)}_{12} = - \frac{e^2 g_Z^2 \,  \beta \, a_{\tau} \, v_{\tau} \, M_Z}{8 \Gamma_Z} \,  \sin^2 (\theta), \nonumber \\
&& R^{(\gamma Z)}_{13} = \frac{e^2  g_Z^2 \, a_{\tau} \, \beta \, M_Z}{8 \, \Gamma_Z} \, \gamma  
\sin(\theta) \, \bigl[ {\rm Re}(Y) - a_{\tau}  {\rm Im}(A) \bigr], 
\nonumber \\
&& R^{(\gamma Z)}_{23} = \frac{e^2  g_Z^2 \, a_{\tau} \, \beta \, M_Z}{8 \, \Gamma_Z} \, \gamma  
\sin(\theta) \, \bigl[ \beta \, {\rm Re}(X) + a_{\tau}  \beta \, {\rm Im}(B)  - v_{\tau} \gamma^{-2} \cos (\theta) \bigr], \nonumber \\
&& R^{(\gamma Z)}_{14} = \frac{e^2  g_Z^2 \, a_{\tau} \, M_Z}{8 \, \Gamma_Z} \, \gamma   \sin(\theta) \, 
\bigl[ a_{\tau} \beta^2 \, {\rm Re}(B) - (1+\gamma^{-2}) {\rm Im}(X) \bigr], \nonumber \\ 
&& R^{(\gamma Z)}_{24} = \frac{e^2  g_Z^2 \, a_{\tau} \, \beta \, M_Z}{8 \, \Gamma_Z} \, \gamma   \sin(\theta) \, 
\bigl[ {\rm Im}(Y) + a_{\tau} (\gamma^{-2} + {\rm Re}(A) \bigr], \nonumber \\ 
&& R^{(\gamma Z)}_{34} = - \frac{e^2  g_Z^2 \, a_{\tau} \, M_Z}{ 4 \, \Gamma_Z} \,  \cos(\theta) \,  {\rm Im}(X), \nonumber   \\
&& R^{(\gamma Z)}_{33} = R^{(\gamma Z)}_{44} = - \frac{e^2  g_Z^2 \, a_{\tau}^2 \, \beta \, M_Z}{4 \, \Gamma_Z} \, \cos(\theta) \, {\rm Im}(A), \nonumber 
\end{eqnarray}
and the remaining components can be obtained using:
\begin{equation}
R_{j,i}^{(\gamma Z)} = R_{i,j}^{(\gamma Z)}  \, \Big|_{X \to X, \, Y \to -Y, \, A \to A, \, B \to -B} 
\quad \text{for} \; \;  i \ne j.
\label{eq:030}
\end{equation}

It is important to identify the components of the matrix $R_{i,j}^{(Z)}$ which are the most sensitive to the weak dipole moments of $\tau$ lepton.
Several observations can be made at this point.
\begin{itemize}
  \item
The terms $R^{(Z)}_{13}, \, R^{(Z)}_{23}, \, R^{(Z)}_{14}, \, R^{(Z)}_{24}$ 
in Eqs.~(\ref{eq:R_Z}) are enhanced due to the large Lorentz factor $\gamma$.
\item
The $\tau$ transverse polarization in the reaction plane along 
the $\hat{x}$ axis is sensitive to real part of $X$. The normal to the reaction 
plane polarization along the $\hat{y}$ axis is sensitive to real part of $Y$. 
And finally, the transverse-longitudinal spin correlation  
$\hat{x} \hat{z}$ and the normal-to-reaction-plane-longitudinal $\hat{y} \hat{z}$ spin correlation 
depend on the imaginary parts of the weak dipole moments.  
Thus, it  may be possible to construct observables sensitive to particular spin components of the $\tau$ leptons.
\item
  We should also note that the interference terms described in Eqs.~(\ref{eq:R_gamma-Z}) are smaller than the $Z$-boson
  contributions by a factor of $\Gamma_Z /M_Z \approx 0.03$, and therefore do not change the dominant pattern.  
\item
Finally,  the longitudinal polarization ${\cal P}_L = r_{34/43}$ of the $\tau^-$ and $\tau^+$ in the SM 
arises due to the vector coupling $v_{\tau}$. This is well known 
and ${\cal P}_L$ was measured already at LEP I, as noted in Ref.~\cite{Mnich:1996hy}. 
The longitudinal polarization, as can be seen from Eqs.~(\ref{eq:R_Z}), 
depends on the weak anomalous magnetic moment ${\rm Re}(X) $ as well.        
\end{itemize}
One can also conclude that the observable studied in Ref.~\cite{Banerjee:2022sgf}, acoplanarity angle between
the planes spanned by the $\tau$ lepton decay products,
is not sensitive to the weak dipole moments at the $Z$-peak. This is because at the $Z$-boson peak,  
other elements of the matrix $R_{i,j}$ rather than the ones contributing significantly at the Belle II energy of 
10.58 GeV~\cite{Banerjee:2022sgf} start to become dominant. 
  
We would like to acknowledge that in case of  $e^- e^+ \to \tau^- \tau^+$ process,
the terms linear in the spins and dipole moments were already derived in 
Refs.~\cite{Bernabeu:1993er, Bernabeu:1994wh}.
However, our results represent some important extensions: they are obtained for various 
initial state fermions $i =(e, \, u, \, d)$, include the complete spin-correlation matrix between the decaying $\tau^-$ and $\tau^+$ leptons, and
also include $Z$-boson exchange, $\gamma$ exchange and $\gamma \, Z$ interference.

\subsection{The  $f_i  f_i \to \tau^- \tau^+$,  $i=\gamma $ case }
\label{subsec:gamma-gamma}

In the description of the reaction $\gamma(k_1) + \gamma(k_2) \to \tau^- (p_-) + \tau^+ (p_+)$ with the 
on-shell photons ($k_1^2=k_2^2=0$),  we separate the contribution of NP from
the total SM contribution to $A(0)$ as described in Ref.~\cite{Eidelman:2007sb}: 
\begin{equation}
A(0)_{SM}= 1.17721(5) \times 10^{-3}, 
\label{eq:005}
\end{equation}
and then define  
\begin{equation}
A(0) = A(0)_{SM} + A(0)_{NP}, \qquad \quad B(0) = B(0)_{NP}.
\label{eq:006}
\end{equation}  
We assume that for the on-shell photons, the dipole moments are real-valued\footnote{We neglect the higher-order contribution to dipole moments coming from the three-photon intermediate state \cite{Bernreuther:2021uqm}, which gives rise to imaginary part of form-factors at $q^2=0$. }.   
  
The matrix element of this reaction is of the order $e^2$ and can be written in terms of spinors of the $\tau$ leptons as: 
\begin{eqnarray}
{\cal M} &=&  \varepsilon_\mu  (k_1) \,  \varepsilon_\nu (k_2) \,  \bar{u} (p_-)  
\Bigl[\Gamma^\mu_\gamma (k_1) \, \frac{\vslash{p}_- -\vslash{k}_1 +m_{\tau}}{t- m^2_{\tau}} \, \Gamma^\nu_\gamma (k_2)  \nn \\
&&+ \Gamma^\nu_\gamma (k_2) \, \frac{\vslash{p}_- -\vslash{k}_2 +m_{\tau} }{u- m^2_{\tau}} \, \Gamma^\mu_\gamma (k_1)  \Bigr]  v (p_+) ,
\label{eq:007}
\end{eqnarray}
where $t = (p_- - k_1)^2$, \ $u=(p_- - k_2)^2$. The four-vectors of the photon polarization, 
$\varepsilon_\mu  (k_1)$ and $\varepsilon_\nu  (k_2)$, obey the conditions: 
$\varepsilon (k_1) \cdot k_1 =\varepsilon (k_2) \cdot k_2 =0$.    

After squaring the matrix element, and averaging over the polarizations of the photons, we obtain
\begin{equation}
|{\cal M}|^2 = R^{{\gamma\gamma}}_{44} + \sum_{i, j=1}^3 \, R^{{\gamma\gamma}}_{i, j} \, s^-_i  s^+_j .
\label{eq:010}
\end{equation}
The elements of the spin-correlation matrix  $R^{\gamma \gamma}_{i, j}/R^{\gamma \gamma}_{44}$ 
depend on the invariant mass of $\tau^- \tau^+ $ pair $s=m_{{\tau^-} {\tau^+}}^2$ and 
the scattering angle $\theta$.  Here, we keep only terms linear in the dipole moments.    

The expressions for the spin-correlation matrix are presented in terms of the 
Lorentz factor $\gamma = \sqrt{s}/(2m_\tau)$ of the $\tau^\pm$ leptons, the velocity 
$\beta$ and the angle $\theta$. 

We define the factor 
\beq
D \equiv  1- \beta^2 \cos^2 \theta. 
\label{eq:011}
\eeq
The elements of the matrix $R^{\gamma \gamma}_{i,j}$ and $R^{\gamma \gamma}_{44}$ are  
(for brevity we denote below $A \equiv A(0)$ and $B \equiv B(0)$):
\begin{eqnarray}
  \label{eq:gamgamR}
R^{{\gamma \gamma}}_{11} &=& \frac{e^4}{8 D^2} \big[ -\beta^2 (\beta^2 - 4 A -2) \cos (4\theta) 
+4 \beta^2 (\beta^2 -2)  \cos(2 \theta)   \nonumber   \\
&& + 4 A \, (7 \beta^2 -8)   - 11 \beta^4 +22 \beta^2 -8 \big],      \label{eq:012}    \\
R^{{\gamma \gamma}}_{12} &=& -R^{\gamma \gamma}_{21} =   
\frac{e^4  B}{4 D^2} \, \beta \, \big( \beta^2 \cos(4 \theta) + 4 \cos(2 \theta) + 15 \beta^2 -20 \big),  \nonumber \\
R^{{\gamma\gamma}}_{13} &=& R^{\gamma \gamma}_{31} =  \frac{e^4 }{2 D^2  } \,  \gamma  \, \beta^2 
\big[ (\beta^2 + A \, (\beta^2-2) -1) \cos(2 \theta) + A \beta^2  - \beta^2  +1 \big] \, 
{\sin(2 \theta)},
\nonumber \\ 
R^{{\gamma \gamma}}_{22} &=& \frac{e^4}{8 D^2 } \big[ -\beta^4 \cos(4 \theta) 
+ 4 \beta^2 (\beta^2 + 4 A) \cos(2 \theta) +16 A \, (\beta^2 -2)  \nonumber \\
 && -11 \beta^4 + 16 \beta^2 -8 \big], \nonumber \\
R^{{\gamma \gamma}}_{23} &=& -R^{\gamma \gamma}_{32} =  \frac{e^4  B}{2 D^2 } \,  \gamma \,  \beta \,
(\beta^2 \cos(2 \theta)  -3 \beta^2 +2) \,  { \sin(2 \theta) },  \nonumber \\
R^{{\gamma \gamma}}_{33} &=& \frac{e^4}{8 D^2} \big[ \beta^2 (\beta^2 - 4 A -2) \cos(4\theta) 
- 4 \beta^4 \cos(2 \theta) +4 A \, (9 \beta^2 -8) \nonumber \\ 
&& + 11\beta^4 + 2 \beta^2 -8 \big], \nonumber \\
R^{{\gamma \gamma}}_{44} &=& \frac{e^4}{8 D^2} \big[ - \beta^4  \cos(4\theta) + 4 \beta^2 ( \beta^2 -4 A -2 ) \cos(2 \theta) - 16 A \, (\beta^2 -2) \nonumber \\
&& - 11 \beta^4 + 8 \beta^2 + 8 \big] .  \nonumber
\end{eqnarray}
{All elements of $R^{\gamma \gamma}_{i, j}$, except the transverse-longitudinal ones,
satisfy the condition $ R^{{\gamma \gamma}}_{i, j} (\theta) =  R^{{\gamma \gamma}}_{i, j} (\pi - \theta)$ for $ 0 \le \theta \le \pi $.}

Note that the contribution from electric dipole moment $B$ is completely separated from the rest of the terms and reads  
\begin{eqnarray}
|{\cal M}|^2_{\rm EDM} &=& \frac{e^4 }{4 D^2} \, \beta \, B \,  
  \Big[ (s^-_1 s^+_2 - s^-_2 s^+_1) (\beta^2 \cos(4 \theta) + 4 \cos(2 \theta) + 15 \beta^2 -20)  \nonumber \\
	&& + 2	(s^-_2 s^+_3 - s^-_3 s^+_2) \, \gamma  (\beta^2 \cos(2\theta) -3 \beta^2 +2) \, \sin(2 \theta) 
	\Big].
\label{eq:013}
\end{eqnarray}
This  can be convenient for studying observables sensitive to $B$.   

Finally, the cross section for the process $\gamma \gamma \to \tau^- \tau^+$ is: 
\begin{equation}
\frac{d \sigma }{d \Omega} (\gamma \, \gamma  \to \tau^- \tau^+) =
\frac{d \sigma }{d \Omega} (\gamma \gamma \to \tau^- \tau^+)\Big|_{{\rm unpol}} 
\, \frac{1}{4}  \Big( 1  +  \sum_{i, j = 1}^3 r^{\gamma \gamma}_{i, j} \, s^-_i  s^+_j  \Big)
\label{eq:014}
\end{equation}
with spin correlation matrix 
$r^{\gamma \gamma}_{i, j} \equiv R^{\gamma \gamma}_{i, j}/R^{\gamma \gamma}_{44}$
and unpolarized cross section
\begin{equation} 
\frac{d \sigma}{d \Omega} (\gamma \gamma  \to \tau^- \tau^+) \Big|_{{\rm unpol}} = \frac{\beta}{16 \pi^2  s } 
\, R^{{\gamma\gamma}}_{44}.
\label{eq:015}
\end{equation}

Note that neither the longitudinal nor the transverse $\tau$ polarization is present for $\gamma \gamma \to \tau^- \tau^+$.
Thus, $r^{\gamma \gamma}_{i,4}=r^{\gamma \gamma}_{4,j}= 0$,  and only the spin-correlation 
part survives in Eq.~(\ref{eq:014}).

\section{Transverse spin correlations of elementary processes}
\label{sec:transverse}

In this section, we present numerical results for $R_{i,j}$ for the process 
$f_i  \, \bar{f}_i \to \tau^- \tau^+$ obtained with analytical formulas of Sec.~\ref{sec:KKMC-th},
as a function of the invariant mass of $\tau$-lepton pair. 

In Figs.~\ref{fig:spin-corr_electron} and~\ref{fig:spin-corr_quarks}, 
we show $r_{11}=R_{11}/R_{44} $ and  $r_{22}=R_{22}/R_{44}$ without including form-factors 
$A(s), \,  B(s), \,  X(s)$ and $Y(s)$,
respectively for $e^-e^+$ and $q \, \bar q$ initial state and two values of scattering angle  $\theta$ = $\pi/3$ and $2\pi/3$.

\begin{figure}
\begin{center}
\includegraphics[scale=0.65]{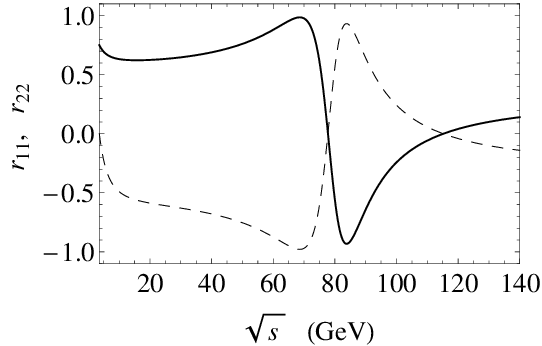}
\includegraphics[scale=0.65]{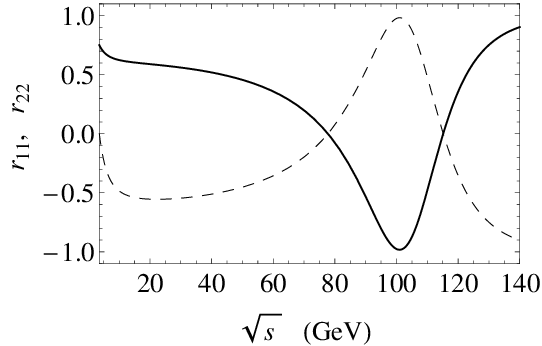}
\includegraphics[scale=0.65]{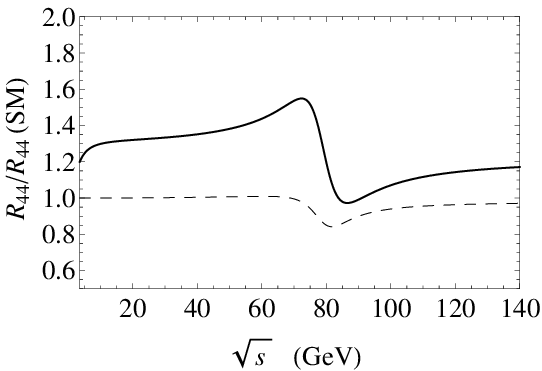}
\includegraphics[scale=0.65]{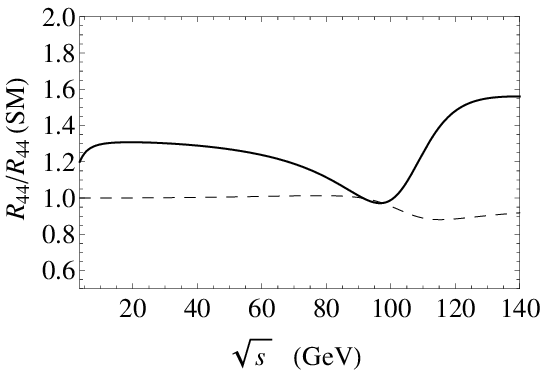}
\end{center}
\centering
\caption{{Upper plots:} transverse spin-correlation elements 
of the $r_{i,j}$ matrix:  $r_{11}$ (solid lines) 
and $r_{22} $ (dashed lines) for the electron-positron initial state. 
The angle $\theta$ is chosen $\pi/3$ for the left plot and $2 \pi/3$ for the right plot. 
The effective  $Z$ couplings to leptons of Table~\ref{tab:Z-couplings} are used.
Dipole moments are not included.
{ Lower plots:  $R_{44}/R_{44}(SM)$ at $\theta=\pi/3$ (left), and 
$\theta=2 \pi/3$ (right). Solid lines:  ${\rm Re}(A(s))=0.1$, dashed lines:  
${\rm Re}(X(s))=0.1$, other form-factors are set to zero.} }
\label{fig:spin-corr_electron}

\begin{center}
\includegraphics[scale=0.65]{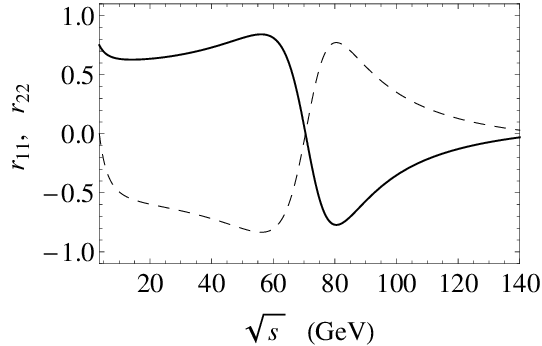}
\includegraphics[scale=0.65]{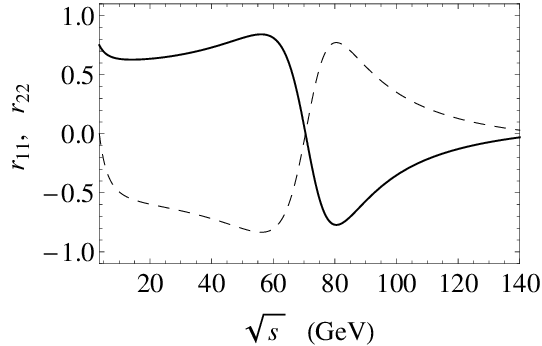}
\includegraphics[scale=0.65]{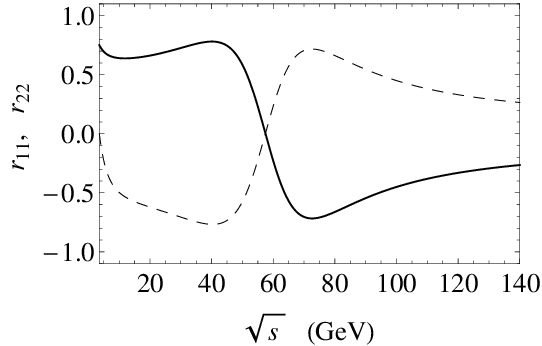}
\includegraphics[scale=0.65]{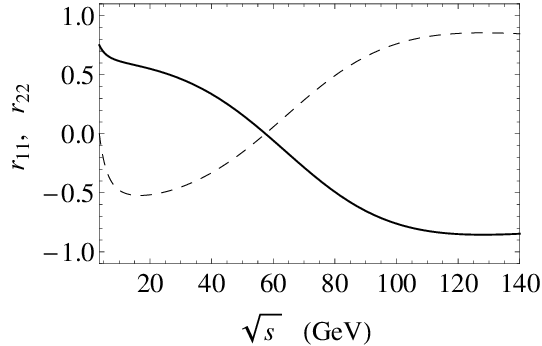}
\end{center}
\centering
\caption{Transverse spin-correlation components $r_{11} $ (solid lines) 
and $r_{22} $ (dashed lines) for the $u \, \bar u$ (top plots) and $d \, \bar d$ (bottom plots) initial states. 
The angle $\theta$ of quark vs $\tau^-$ is chosen $\pi/3$ in the left plots and $2 \pi/3$ in the right plots. 
The effective couplings of $Z$ to quarks from Table~\ref{tab:Z-couplings} are used. Dipole moments are not included.}
\label{fig:spin-corr_quarks}
\end{figure}

\begin{figure}[!htb]
\begin{center}
\includegraphics[scale=0.65]{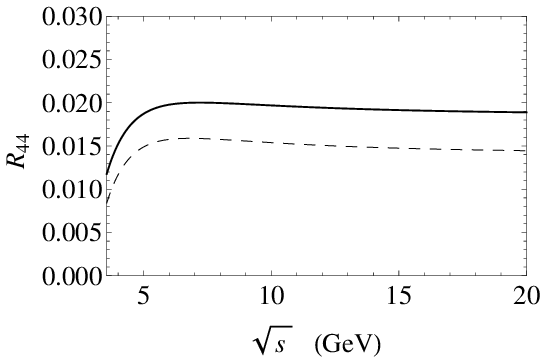}\\
\includegraphics[scale=0.65]{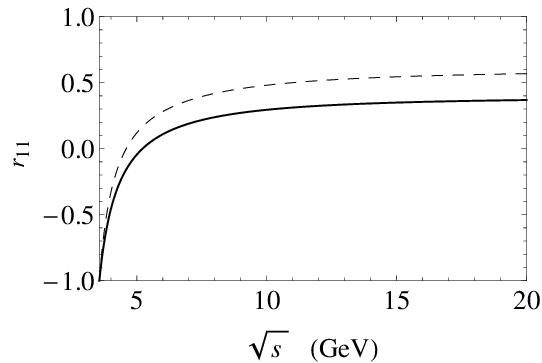}
\includegraphics[scale=0.65]{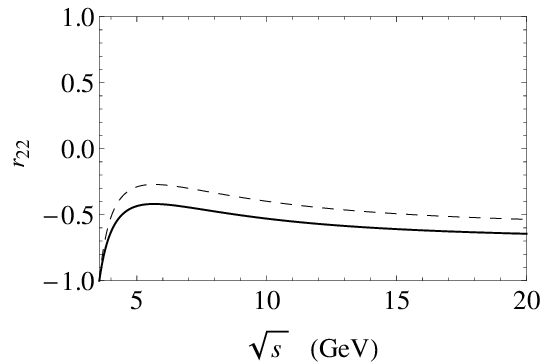}
\includegraphics[scale=0.65]{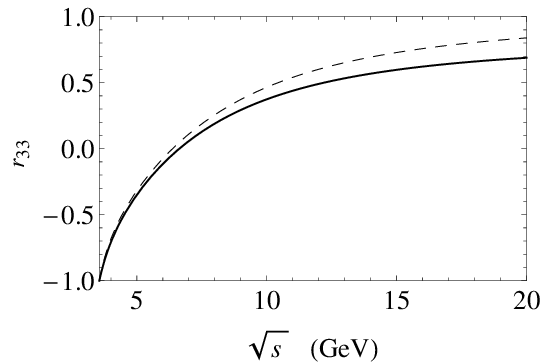}
\includegraphics[scale=0.65]{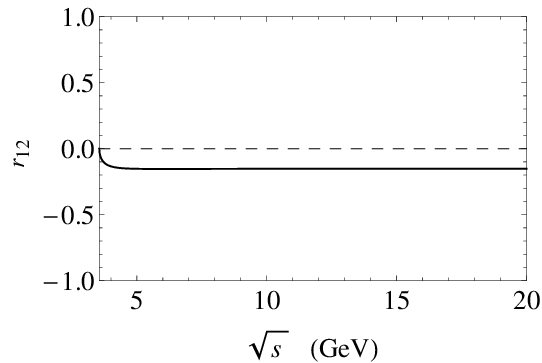}
\includegraphics[scale=0.65]{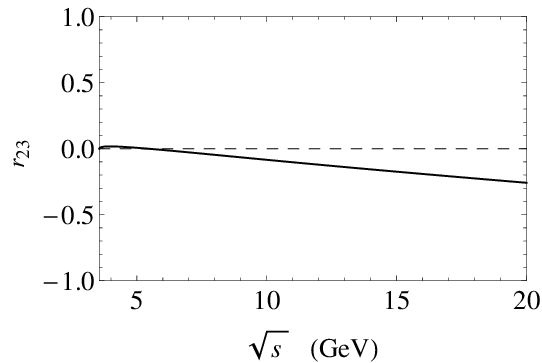}
\includegraphics[scale=0.65]{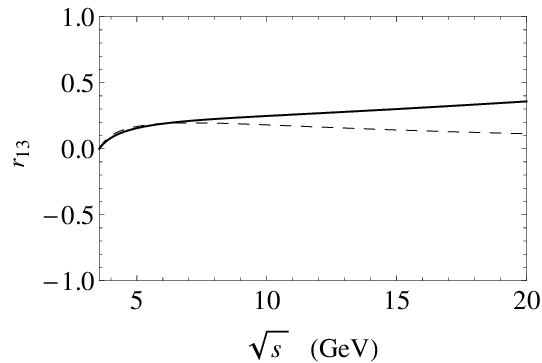}
\end{center}
\centering
\caption{Energy dependence of  $R^{\gamma \gamma}_{44}$ and 
$r^{\gamma \gamma}_{i,j} $.  
Solid lines are calculated for the values $A=0.1$ and $B=0.1$;  dashed lines -- for $A=B=0$. 
The angle $\theta$ is chosen $\pi/3$. }
\label{fig:spin-corr_gamma-gamma}
\end{figure}

We also used distributions from Fig.~\ref{fig:spin-corr_electron} to test our formulas with respect to the predictions of {\tt KKMC},
including test of the frames conventions. The plots of other elements $r_{i,j}$ are not as interesting, because they are very small (except for the $\tau$ mass effects), and therefore are not shown.   

The pattern of the transverse spin correlations, $r_{11}$ and $r_{22}$, is mostly dominated by the SM correlations,
and is not straightforward to explain. It depends on the flavor and scattering angle, and as a consequence,
in case of $q \bar q$ initial state, it is parton distribution function (PDF) dependent.
The electron-positron case is more straightforward in terms of definition of a suitable observable, as the incoming state is always $e^-e^+$. 

The transverse spin correlations for $e^- e^+$ and $u \bar{u}$ change signs at energies below and above the $Z$-peak, and then for higher energies stabilize at small positive value, except for $e^- e^+ $ at $\theta = 2 \pi/3$, where value is larger. 
This is not the case for $d \bar d$ initial state, where correlations change sign only once below Z-boson peak, 
and then remain negative and small (large) for $\theta = \pi/3$ ($\theta = \pi/3$). 
Above the $Z$-boson peak, they  only very weakly  depend on energy.

{In addition, in Fig.~\ref{fig:spin-corr_electron} (lower plots) we show 
dipole form-factors impact on the  $R_{44}$ which determines the cross section 
of $e^-e^+ \to \tau^- \tau^+$ of unpolarized $\tau$ leptons. The ratios of $R_{44}$ calculated with
the form-factors ${\rm Re}(A(s))$ and ${\rm Re}(X(s))$ included, to $R_{44}(SM)$ calculated without 
form-factors, are presented.  As is seen, there is a clear effect of the real parts of anomalous magnetic 
and anomalous weak magnetic form-factors.  The electric form-factors $B(s)$ and $Y(s)$ 
do not contribute to $R_{44}$, while the contribution from ${\rm Im}(A(s))$ and ${\rm Im}(X(s))$ 
is very small and is not shown.} 
 
In Fig.~\ref{fig:spin-corr_gamma-gamma}, we show $R^{\gamma \gamma}_{44}$
and other elements $r^{\gamma \gamma}_{i,j}$ for the process $\gamma \gamma \to \tau^- \tau^+$ with and without dipole moments included.   
The chosen values of dipole moments $A(0)=0.1$ and $B(0)=0.1$ are perhaps unrealistically large, but allow us to demonstrate
the sensitivity of various elements to the dipole moments.  In particular, one can observe that the elements 
$r^{\gamma \gamma}_{12}$ and $r^{\gamma \gamma}_{23}$ vanish in the absence of the electric dipole moment, 
and become finite for nonzero values of it, while the other elements depend only on the anomalous magnetic dipole moment.   

\section{Reweighting procedure}
\label{sec:reweighting}

The analytical formulas presented in Sec.~\ref{sec:KKMC-th} have been implemented into 
reweighting algorithms for {\tt KKMC} and {\tt TauSpinner} programs.
Necessary tests were performed to confirm that event weights at fixed $\tau$-pair 
virtuality and incoming fermion flavor reproduce  those formulas correctly.

\subsection{The reweighting algorithm for {\tt  KKMC}: case of  anomalous dipole moments}
\label{sec:algorithm1}

With respect to implementation prepared in Ref.~\cite{Banerjee:2022sgf},
the function calculating the spin correlations and an overall event weight
has been updated to include contributions due to $Z$-boson exchange and $Z\gamma$-interference.  
The formulas have become substantially longer after introducing the anomalous couplings of the $\tau$ leptons 
to the $Z$-boson along with coupling to the virtual photon. Other technical aspects explained in
Ref.~\cite{Banerjee:2022sgf} remain valid.

However, for higher energies the $p_T$ of the intermediate boson can become larger than $m_\tau$
because of more frequent initial state bremsstrahlung.
Thus, the validity tests of improved Born approximation for constructing the event weight
should be repeated. The usage of the reference frames as  described
in Ref.~\cite{Richter-Was:2020jlt,Richter-Was:2016mal} may be needed.
This is true both for
$e^-e^+$ and $p p$ applications, as discussed in the following section.
We will return to this point in a future work, when important details
of planned experiments become available and impact of bremsstrahlung on the analysis procedure
can be quantified.

Another minor detail that also has been improved in the present version of the code is an explicit rotation
by angle $\pi$ around $y$-axis, which is necessary to match the reference frame used in our present
formulas and in the {\tt KKMC} code.

\subsection{The reweighting algorithm for {\tt TauSpinner} }
\label{sec:algorithm}

The basic formalism  of {\tt TauSpinner} is documented in Eqs.~(7)--(12) in Sec.~2.2 
of Ref.~\cite{Przedzinski:2018ett}. 
Here we assume that reader is familiar with this reference,
and on the details of how the kinematics of hard process can be deciphered from the information available in the event record.

Let us briefly recall the basic  equation used in the reweighting algorithm:
 \begin{eqnarray}
&d \sigma = \sum_{flavors}  dx_1 \, dx_2 \, f(x_1,...)\, f(x_2,...) \, d\Omega^{parton\; level}_{prod} \; d\Omega_{\tau^+} \; d\Omega_{\tau^-} \nonumber \\
& \times \Bigl(\sum_{\lambda_1,  \lambda_2 }|{\cal M}^{prod}_{parton\; level}|^2 \Bigr)
 \Bigl(\sum_{\lambda_1 }|{\cal M}^{\tau^+}|^2 \Bigr)
 \Bigl(\sum_{\lambda_2 }|{\cal M}^{\tau^-}|^2 \Bigr) \, wt_{spin},
\label{eq:parton-level}
\end{eqnarray}
which represents product of phase-space integration elements and matrix elements, squared and averaged over spin. Sum over flavors of incoming
 partons and integration over PDFs is also given. Let us provide some details.
 The sum over flavors is followed by integration elements over incoming partons energy fractions and PDFs.
 Then, the Lorentz invariant phase-space integration elements at the parton level to $\tau$-pair production and for the decay of $\tau^+$ and $\tau^-$ follow.
 Terms in  brackets correspond to spin averaged matrix elements for $\tau$ production and each $\tau$ decay respectively.  Sum over $\tau$-decay channels is not given explicitly. 
 Only the  spin weight
\begin{equation}
wt_{spin} = \sum_{i ,j=t,x,y,z} r_{i, j} h^i_{\tau^+} h^j_{\tau^-} \label{eq:wtspin}
\end{equation}
depends on kinematics of both $\tau^\pm$ production and decay in a rather simple way, as explained below. 
It is a smooth function, which in addition satisfies the conditions 
$ \left\langle wt_{spin} \right\rangle =1 $ and $0<wt_{spin}<4 $.

In Eq.~(\ref{eq:wtspin}), $h^i_{\tau^+} $ and $ h^j_{\tau^-}$ denote the 
polarimetric vectors, which for a particular $\tau^+$ or $\tau^-$ decay channel depend on the kinematics of $\tau^\pm$ decays. The $r_{i, j}$ 
matrix combining each $\tau$ spin-polarization and $\tau$-pair spin-correlation matrix
depends on $\tau$-production kinematic (including incoming parton flavors) only.

For reweighting, the ratio of expressions given by Eq.~(\ref{eq:parton-level}) is calculated for two distinct
assumptions of the matrix
elements. Obviously, the phase-space integration elements cancel out in the ratio. For the sums, several rather trivial options are possible.
In particular, they depend on which terms in the matrix element used in event sample generation differs from the one used in the variant to be implemented.

In this update, starting with the $q \bar q \to \tau^- \tau^+$ parton level  processes,
the functions calculating the spin correlations and an overall event weight
have been updated in the {\tt TauSpinner} with analytical formulas of Sec.~\ref{sec:KKMC-th}.
Before this update, the transverse spin correlations in process $q \bar{q} \to Z/\gamma \to \tau^- \tau^+$ were available via the
the EW library SANC~\cite{Andonov:2004hi,Arbuzov:2020coe} and its predefined tables, which were neither very flexible nor transparent.   

Next, let us turn our attention to the new  parton level process:
$\gamma \gamma \to \tau^-\tau^+$.  The formulas for spin amplitudes, cross section and spin-correlation
matrix were presented in Sec.~\ref{sec:KKMC-th}.
Eqs.~(\ref{eq:012}) contain all necessary expressions.
It is important to note that each parton process contributes incoherently to the final
state. Thus, the introduction of nearly on-shell photon as an incoming parton, 
and a corresponding hard process, is possible by straightforward extension  
of the sums in Eq.~(\ref{eq:parton-level}) only.

For the set of PDFs, we can take the ones described in Refs.~\cite{Klein:2016yzr, Xie:2021equ}.
These structure functions include the photon PDF as well, but this choice may be not straightforward.
The actual choice may depend on the details of the selection criteria of the heavy ion collision events. In some cases, the
contribution from the $\gamma$ PDFs may even dominate, and the usually large Drell-Yan contributions may become relatively small.

In terms of technical details of the {\tt TauSpinner}, we again assume that the reader is familiar with the ones described in Ref.~\cite{Przedzinski:2018ett}.
Minor extensions to include additional parton level $\gamma\gamma$ process has been introduced in this update\footnote{
Practical comments can be found in the code of the latest version of  {\tt TauSpinner}, downloadable from the project 
web page {\tt http://tauolapp.web.cern.ch/tauolapp/}.}.

Three new functions have been  added to the {\tt TauSpinner} code, and the sum in function 
{\tt double sigborn(...)} was extended to the $\gamma\gamma$ process.
In the function  {\tt dipolgammarij(...)}, we calculate the  $R_{i,j}^{\gamma \gamma}$ matrix.
The functions \\
{\tt T\_gamm(MODE,SVAR,COSTHE,TA,TB)}  and  {\tt T\_gammNEW(MODE,SVAR,COSTHE,TA,TB)}
calculate cross section normalized as in older parton level processes;  anomalous dipole moments 
are absent in {\tt T\_gamm() }.
  The function  {\tt T\_gamm}  can be set to return zero
with  internal parameters of the algorithm. This feature is useful
for reweighting samples which have no $\gamma \gamma$ contribution.
The functions {\tt T\_gamm} and {\tt T\_gammNEW}  are called by the algorithm with input parameters named
{\tt MODE, SVAR, COSTHE, TA,} and {\tt TB}
respectively. The  {\tt MODE} 
is not used at present. The Mandelstam $s$ variable {\tt SVAR}, cosine of the hard process scattering angle
{\tt COSTHE} and helicities/chiralities {\tt TA,} and {\tt TB} of the outgoing $\tau$'s are set internally
based on the event kinematics, and are not expected to be initialized by the user.

The {\tt T\_gamm} function does not depend on anomalous moments, which are introduced via call to {\tt T\_gammNEW} only, and subsequent call to {\tt dipolgammarij}.
The  parameters (constants) which require initialization by the user in the appropriate {\tt struct}  or common blocks are:

  \begin{itemize}
    \item {\tt  IFgammOLD}: allows  to switch $\gamma \gamma$ contribution on/off  in weight denominator
      according to what is present in generated sample,
     \item {\tt  Adip}: anomalous magnetic moment,
     \item {\tt  Bdip}: anomalous electric moment,
     \item {\tt  iqed=iqedDip}: 
       {to include  SM anomalous magnetic dipole moment in $\gamma \gamma$ contribution
         in the denominator of the ratio  defining weight, see  Eqs. (19), (20) of Ref.~\cite{Banerjee:2022sgf}. The choice depends on whether
         the contribution is taken into account in the sample to be reweighted.
       Note that this contribution is always present in the weight in the numerator, unless photon structure function is absent (set to zero).}
  \end{itemize}
  
The implementation of the initializations parameters currently hard-coded in routine {\tt T\_gamm} is not universal.
The actual choice of parameters can be generalized following discussions with interested users.

As is the case with  all parton level processes of Eq. (\ref{eq:parton-level}) presented in Ref.~\cite{Przedzinski:2018ett}, the $\gamma\gamma$ functions  {\tt T\_gamm} and {\tt T\_gammNEW} provide results with 
$\tau$ leptons  helicity-level approximation. The
transverse spin correlations are not included at this step in the algorithm.  Complete spin correlations are
introduced later in the algorithm step while calculating and using $wt_{spin}$.
From now, in the loop over summation used to calculate the averaged $R_{i, j}$ matrix on the basis of Eq.~(\ref{eq:parton-level}),
contributions of $R_{i, j}$  from $\gamma\gamma$,  Eqs.~(\ref{eq:gamgamR}), are introduced. 

The implementation can be  adapted for reweighting events of  ultra-peripheral heavy-ion collisions,
where it can be used for the analysis of $\tau$ dipole moments at the LHC, 
as done in the Refs.~\cite{ATLAS:2022ryk,CMS:2022arf}.

\section{Numerical results for semi-realistic observable}
\label{sec:numerical}

In Ref.~\cite{Banerjee:2022sgf}, we discussed the observables
sensitive to  contribution of the dipole form-factors in the $ e^- e^+ \to \gamma^* \to \tau^-\tau^+$ 
interaction at Belle II energies.
We found that transverse spin correlations of the $\tau$-pair production,
combined with $\tau^\pm \to \pi^\pm \pi^0 \nu_\tau$ decays, may be useful for
that purpose. For these energies, the transverse spin correlations, calculated at
the lowest order in the SM are simple, and weakly dependent on the c.m. energy. Transverse spin
correlations in the direction perpendicular to (aligned in) the reaction plane
are of the opposite sign. Therefore, fortunately the corresponding kinematic configurations
are easy to separate,  and the initial-state bremsstrahlung (ISR)
emissions do not contribute in sizable amounts to the transverse momenta of 
the $e^-e^+ \to \tau^- \tau^+ n \gamma$ events. The transverse momenta of unobservable photons  $p_T^{ISR}$
are much smaller than $m_\tau$ and, as a consequence,
also smaller than the transverse momenta of the $\tau$-decay products.

\subsection{Observables sensitive to dipole moments at $Z$-boson pole}
\label{subsec:observables Z}

At the $Z$-boson peak, the spin-correlation pattern is different from that at low energies
relevant for the Belle II experiments. The largest components of the $R$ matrix are of longitudinal-transverse type, e.g. the terms $R_{13}$, $R_{31}$,$R_{23}$ and $R_{32}$ are large.
That is why the observable of Ref.~\cite{Banerjee:2022sgf} is not sensitive to the anomalous magnetic and electric dipole form-factors.

We have chosen therefore different and also rather simple observable, for the case when
both $\tau$ leptons decay to a $\pi$ and a $\nu_\tau$.
In the center of mass of $\tau^- \tau^+$ pair, we take the energy to correspond to the mass of $Z$-boson.
That should, in realistic conditions, reduce the size of the initial-state
bremsstrahlung, even though the width of the $Z$-boson is larger than the mass of the $\tau$-lepton.
This aspect will require large care and detailed understanding of detector conditions
in the future experiments at the Future Circular Collider (FCC-ee) at the $Z$-boson peak.

We calculate vector products of the particle momenta
\begin{equation}
\vec{v}_1 = \vec{p} \times \vec{k},  \quad 
\vec{v}_2 = \vec{p}_{\pi^-} \times \vec{p}_{\nu_\tau}, \quad 
\vec{v}_3 =  \vec{p} \times \vec{v}_1, 
\label{eq:031}
\end{equation}
and normalize these three-vectors to the unit length: $\hat{v}_i = \vec{v}_i / |\vec{v}_i|$ 
($i=1,2,3$).
Here $\vec{p}_{\pi^-}$ and $\vec{p}_{\nu_\tau}$ are the momenta of $\pi^-$ and $\nu_\tau$
coming from the decay $\tau^- \to \pi^- \nu_\tau$.  Then the acoplanarity angle between the plane 
spanned on  the vectors $\vec{k}$
and $\vec{p}$,  
and the plane spanned on the vectors $\vec{p}_{\pi^-}$ and $\vec{p}_{\nu_\tau}$, is determined from the relations
\begin{equation}
\cos (\varphi)= \hat{v}_1 \cdot \hat{v}_2, \qquad  \sin(\varphi)= \hat{v}_2 \cdot \hat{v}_3.
\label{eq:032}
\end{equation}
This acoplanarity angle spans the range from $0$ to $2 \pi$. 

The number of events with and without dipole moments included vs acoplanarity angle is calculated for the energy, 
corresponding to the $Z$-boson peak $\sqrt{s} = M_Z=91.1876$ GeV. In these conditions the $Z$ exchange
plays the dominant role, and the $\gamma Z$ interference gives a minor contribution.
Therefore it becomes possible to 
investigate effect of the weak dipole moments $X(M_Z^2)$ and $Y(M_Z^2)$. 
It is rather clear from Eqs.~(\ref{eq:R_Z}) that the dominant dipole moment
contributions $R^{(Z)}_{13}, \, R^{(Z)}_{23}, \, R^{(Z)}_{14}, \, R^{(Z)}_{24}$ are proportional 
to $\cos(\theta)$ and therefore would vanish if integrated over the whole region of $\theta$.   
However, assuming that the $\tau$ momenta are reconstructed, we can select the forward or backward ranges of the angle $\theta$, and thus, obtain sensitivity to the dipole moments.     
Results of the calculations are presented in Fig.~\ref{fig:acoplanarity}. 
      
\begin{figure}[!htb]
\begin{center}
\includegraphics[scale=0.35]{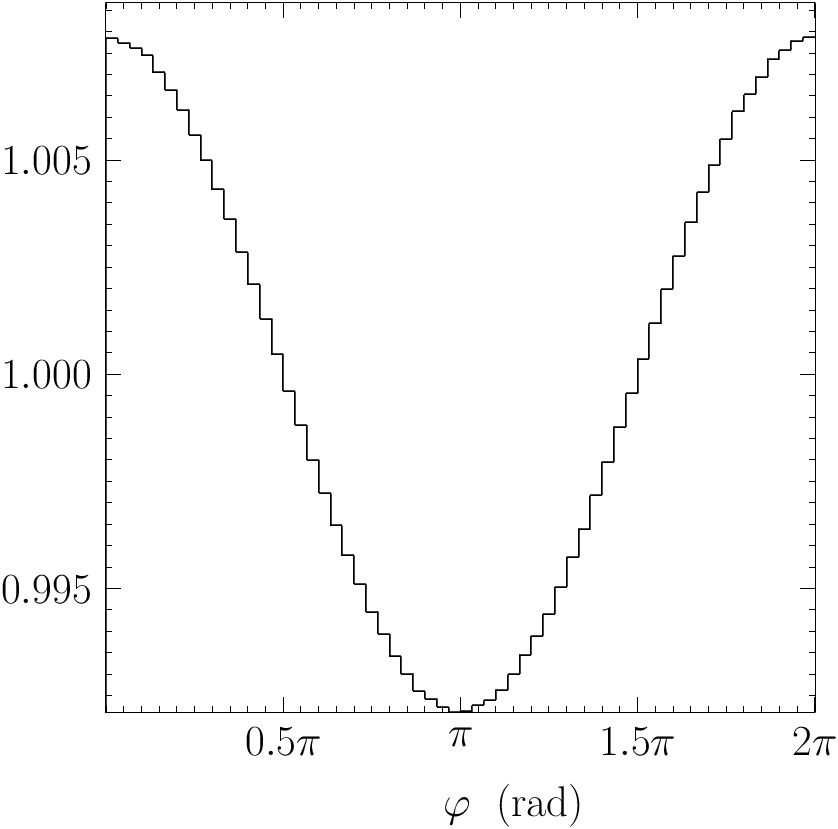}
\includegraphics[scale=0.35]{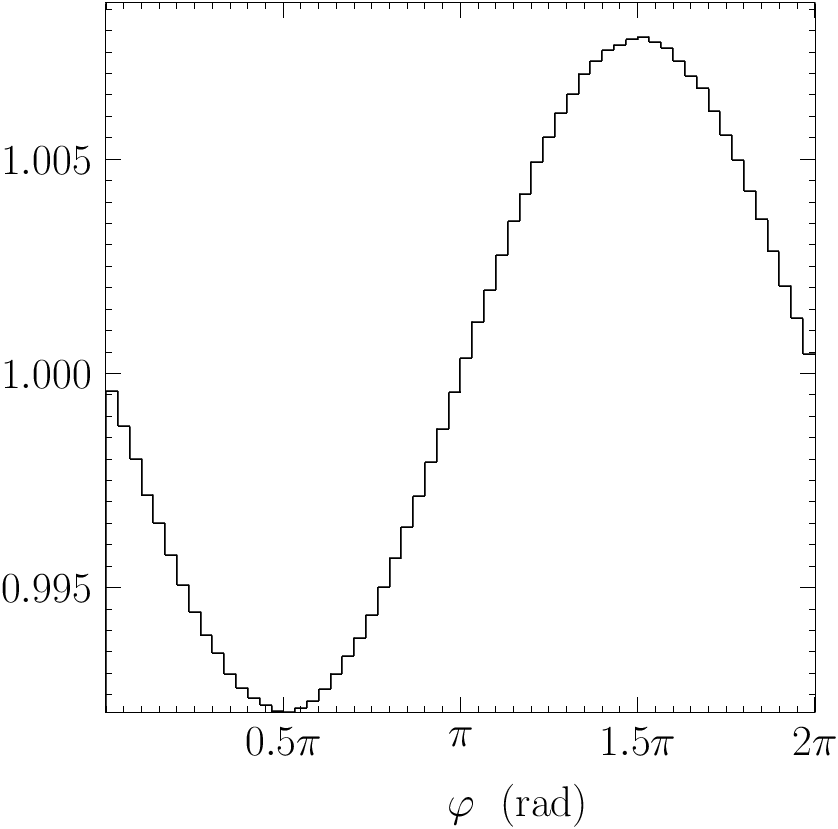}
\includegraphics[scale=0.35]{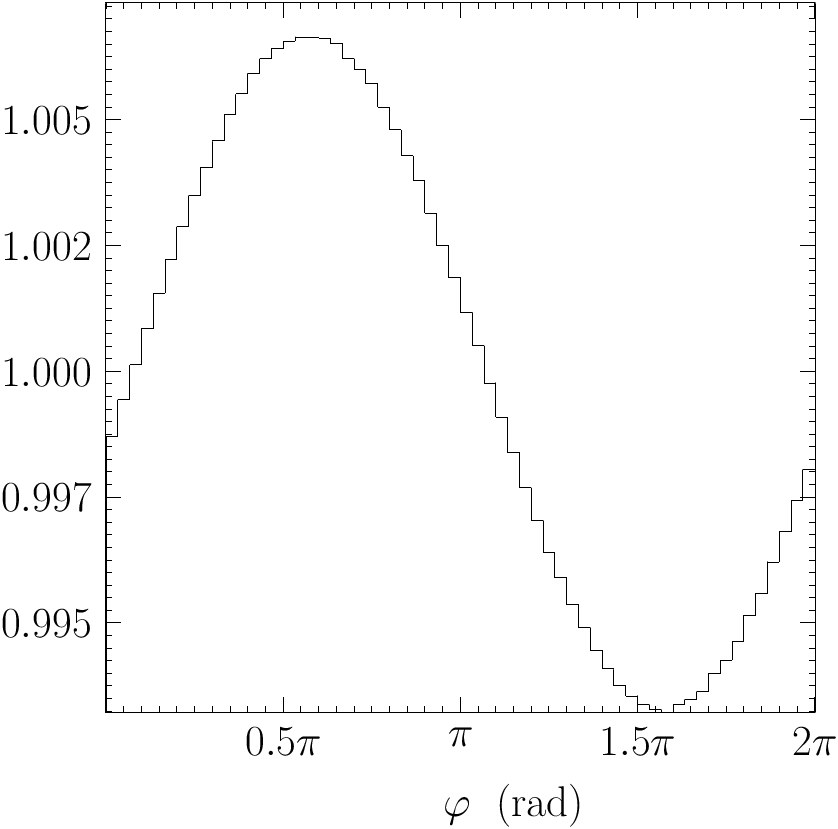}
\includegraphics[scale=0.35]{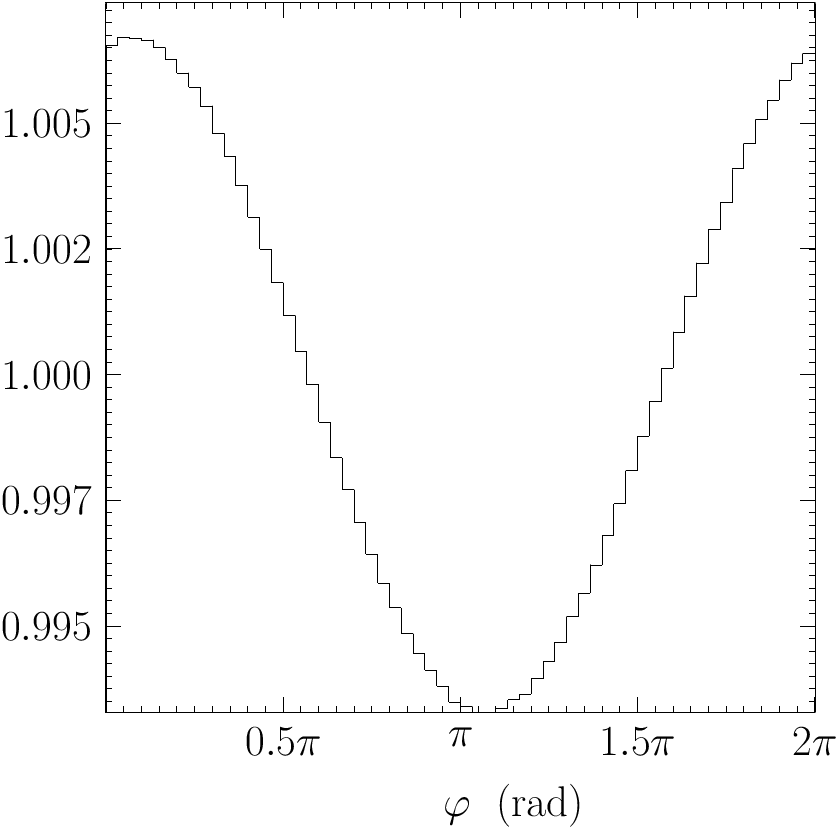}
\end{center}
\centering
\caption{Ratio of number of events with and without weak dipole moments included, 
as a function of the acoplanarity angle $\varphi$ at the energy $\sqrt{s}=M_Z$.  
The backward angles $\theta$ are selected,  i.e. $\cos(\theta)<0$. 
The top left plot is calculated for ${\rm Re}(X) = 0.0004$, the top right plot is calculated for ${\rm Re}(Y) = 0.0004$, 
the bottom left is calculated  for  ${\rm Im}(X) = 0.0004$, and the bottom right is calculated  for ${\rm Im}(Y) = 0.0004$. 
For the imaginary form-factors, the constraint  $E_{\pi^+} > E_{\bar{\nu}_\tau}$  
is applied on the $\tau^+$ side.   
The magnetic and electric form-factors $A(M_Z^2)=B(M_Z^2)=0$ are used.}
\label{fig:acoplanarity}
\end{figure}

As it is seen from Fig.~\ref{fig:acoplanarity} (top), the real part of   
$X(M_Z^2)$ and $Y(M_Z^2)$ generate, respectively, 
the event distribution $\sim 1 + 0.008 \cos( \varphi)$ and  $\sim 1 - 0.008 \sin(\varphi) $. 
As pointed in Subsec.~\ref{subsec:e-e_q-qbar} in discussion of Eqs.~(\ref{eq:R_Z}), these observables are chosen
to be sensitive to the transverse and normal to the reaction plane $\tau$ spin degree of freedom. 

The imaginary parts of $X(M_Z^2)$ and  $Y(M_Z^2)$ result in the event distributions 
shown in Fig.~\ref{fig:acoplanarity} (bottom). Since they involve transverse-longitudinal, $R_{13}$, and 
normal-to-reaction-plane-longitudinal, $R_{23}$,
spin correlations, these distributions appear sensitive to dipole moments only if  the longitudinal component of the momentum of $\pi^+$,   
which comes from the $\tau^+ \to \pi^+ \bar{\nu}_\tau$ decay,  
is constrained.
To increase the magnitude of the distributions,
we impose the condition $E_{\pi^+} > E_{\bar{\nu}_\tau}$ on the energies of $\pi^+$ and $\bar{\nu}_\tau$. 

As one can see, the imaginary part of $X(M_Z^2)$ and of $Y(M_Z^2)$ generates, respectively, 
the distribution $\sim 1 + 0.008 \sin(\varphi - \delta)$ and  $\sim 1 + 0.008 \cos(\varphi - \delta^\prime)$ with small 
shifts $\delta$ and $\delta^\prime$. 
These shifts are caused by non-dominant spin-correlation coefficients, for example, 
by $R_{12}^{(Z)}$, which slightly modifies distribution for ${\rm Im}(X(M_Z^2))$. 
    
The magnitude of the weak dipole moment effects in all distributions is about 0.008. 
This may seem too large in view of the chosen very small value 0.0004 of the dipole moments. The 
observed enhancement is due to the large Lorentz factor at the $Z$-peak.

\subsection{The effect of $Z$-boson exchange at Belle II energy }
\label{subsec:observables Belle}

Let us turn our attention to lower energies. In Ref.~\cite{Banerjee:2022sgf}, we studied effects of the dipole
form-factors $A(s)$ and $B(s)$ on spin effects and on an observable of potential interest for the Belle II measurement.
Since this is expected to be a precision measurement,  that is why the effect of $Z$-boson exchange, or better
to say the $Z \gamma$ interference, which could represent potential bias, should be addressed. 
In Fig.~\ref{fig:Belle-energy}, we present the ratio of distributions where the $Z$-boson contribution is turned on or off. The distributions are prepared exactly in the same 
manner as for Fig.~2 of Ref.~\cite{Banerjee:2022sgf}. The $\tau^\pm \to \pi^\pm \pi^0 \nu_\tau$ decay channels are used for the production process $e^-e^+\to \tau^-\tau^+ n\gamma$ at Belle II energy.
  Acoplanarity between the planes spanned by the $\pi^-\pi^0$ system (from $\tau^-$)   and the $\pi^+\pi^0$ system (from $\tau^+$), respectively, 
   is monitored. The momenta are boosted to the rest frame of visible decay product system
  of $\pi^-\pi^0\pi^+\pi^0$ mesons. An additional selection criteria
  on the pions energies, $(E_{\pi^-}-E_{\pi^0})(E_{\pi^+}-E_{\pi^0})$ to be larger or smaller than zero, is applied.

The anomalous magnetic and electric form-factors are not taken into account. We see that the bias on the spin-observable sensitive to dipole moments 
due to $Z$-contributions is at the level of $10^{-5}$. Thus, the effect of $Z$-boson exchange can be considered
to give a rather small correction even when analyzing $> 10^{10}$ events that are expected to be collected at the Belle II experiment.
We can conclude that  the  observable of Ref.~\cite{Banerjee:2022sgf} remains useful  for studying the magnetic and electric dipole form-factors at the Belle II energies.

\begin{figure}[!htb]
\begin{center}
\includegraphics[scale=0.55]{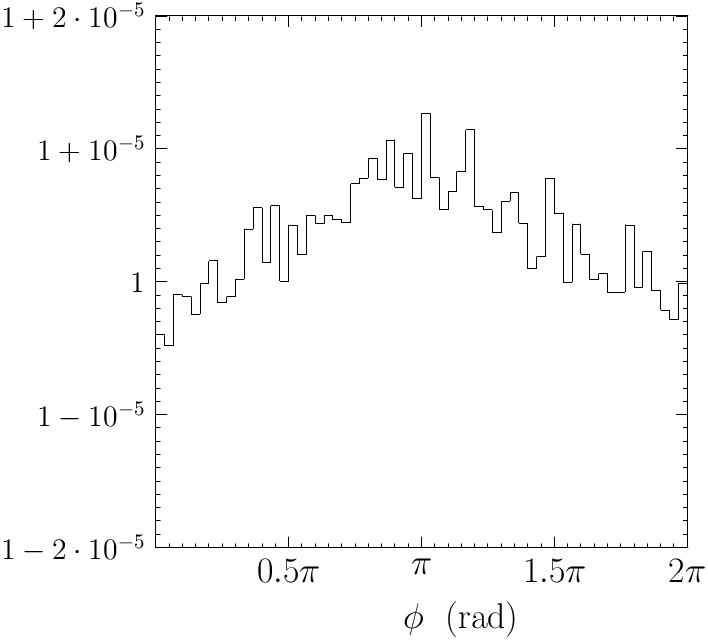}
\end{center}
\centering
\caption{Impact of $Z$-exchange to the acoplanarity of  $\pi^\pm \pi^0$
  planes of $\tau^\pm$ decay products. The production process  $e^+e^- \to \tau^+\tau^-n\gamma$,
  and decays $\tau^\pm \to \pi^\pm \pi^0 \nu_\tau$ at the c.m. energy 
  $10.58$ GeV are used.   Events are selected so that there is
  the same sign of energy differences for the charged and neutral pions coming from decays  
	$\tau^- \to \pi^-  \pi^0 \nu_\tau$ and $\tau^+ \to \pi^+ \pi^0 \bar{\nu}_\tau$.
	The detailed definition of presented observable is given
  in Ref.~\cite{Banerjee:2022sgf}. The ratio of the acoplanarity
  distributions with and without $Z$-boson contribution is shown. }
\label{fig:Belle-energy}
\end{figure}

On the other hand, we find that  the observable of Ref.~\cite{Banerjee:2022sgf} does not have much sensitivity to the 
dipole moments for  higher energies, of the  $Z$-boson peak. In this case, an alternative observable is needed,
possibly the one presented in the previous subsection.

\subsection{ The case of $pp$ collisions}

Constructing observable sensitive to dipole moments in the $pp$ collisions is far more difficult.
The signatures of anomalous dipole moments in $\gamma \gamma$ parton level collision can not be separated
from parton processes of $q \bar q$ collisions.  Also, contrary to the
$e^+e^-$ case, the choice of the direction of the $\hat{z}$ axis in the reference frame to be used
can be ambiguous. Finally, defining the $p_T$ of the hard reaction system does not seem to be straightforward.

In $\gamma \gamma \to \tau \tau$ process, the largest sensitivity to the dipole moments come from
$R^{\gamma \gamma}_{13}, \, R^{\gamma \gamma}_{31}$ 
and $R^{\gamma \gamma}_{23}, \, R^{\gamma \gamma}_{32}$ elements of $R^{\gamma \gamma}_{i,j}$ matrix.
This imply that our choice of the $\tau$ decays could be
$\tau^\pm \to \pi^\pm \nu_\tau$ and $\tau^\mp \to \pi^\mp \pi^0 \nu_\tau$.
The observable could be the difference of the $\pi^\pm$ energy spectra
in sub-samples defined by whether difference of the $\pi^\mp$ and  $\pi^0$
momenta tend to be closer to reaction plane or outside.
The backgrounds, e.g. the Drell-Yan quark level processes, do not have large transverse-longitudinal 
$R$-matrix components. At low virtualities, the  SM contribution to Drell Yan 
$R^{(\gamma) }_{23/32}$ is zero, and in the relativistic limit $R^{(\gamma)}_{13/31}$
is non-dominant. 
Also in peripheral collisions, the contribution from the Drell-Yan 
should be reduced. This impact of spin correlations on dipole moment phenomenology offers an interesting
starting point for further work with more details of hadron collider conditions taken into account.

\section{Summary and outlook}
\label{sec:summary}

In this paper, we have discussed effects from  electric and anomalous magnetic dipole moments of the $\tau$ leptons
on transverse spin correlations in the $\tau$ pair production and decay.

These studies include calculation of the analytical formulas for spin-amplitude components
and spin-correlations matrices, for $e^- e^+ \to \tau^- \tau^+$,  $q \bar q \to \tau^- \tau^+$
and   $\gamma \gamma \to \tau^- \tau^+$ processes. In case of  the $s$-channel  $f_i \, \bar f_i \to \tau^- \tau^+$ process,
including photon and $Z$-boson exchange, their interference  
{ and radiative corrections in framework of IBA}, makes those formulas applicable
for a large range of $\tau$-pair invariant mass,
from the Belle II energies {up to above}  $WW$ and $ZZ$ { pair production} thresholds.
Numerical predictions for the ratios of transverse components of spin-correlations matrices
were also shown and discussed.

The analytical formulas are embedded into algorithms for generated events reweighting,
to be used with {\tt KKMC} generator for events produced in $e^-e^+$ collisions and {\tt TauSpinner}
program for $pp$ collisions.

To demonstrate applications of the reweighting algorithm, semi-realistic observables are suggested, and
for  $e^-e^+$ case numerical results are presented for anomalous dipole moments at high energies and for
$Z$-boson exchange contribution  to Belle II energies.

\vspace{0.2cm}
\centerline{\it Technical developments}

The algorithm for the implementation of anomalous magnetic and electric dipole moments for the s-channel photon
exchange in $e^+e^- \to \tau^+\tau^- $ process
was enriched with contribution of $Z$-boson exchange. With the help of the extension to include
$Z$-boson contributions, the impact on dipole moment observable ambiguities at Belle II energies can be studied. The algorithm
can be used also for higher energies in its present form up to {and above} $ZZ$-boson pair production threshold.
{
{ At these energies,  contributions from $WW$- and $ZZ$-boson pair production become double resonant 
and box diagram contributions become sizable. Their impact on spin effects cannot be neglected.}   

The observable, acoplanarity of decay products of $\tau$ lepton pairs,
survive upgrade to high energies after minor modifications, provided the effects of photon radiation
can be ignored{, as in the case of $\sqrt{s} \simeq M_Z^2$}. We have checked the results for a new observable, and also  with the help of analytic calculations. We can conclude that  tools for further studies
of more experimental details is prepared. It is  again an add-up to
{\tt KKMC}  and may be useful not only for the FCC oriented studies, but also for the evaluation of potential biases
in the Belle II precision measurements as well.

We have prepared and installed in {\tt TauSpinner} preliminary version of the code necessary for including
$\gamma \gamma \to \tau^- \tau^+$ parton level process in event reweighting. It can be added with or without
including anomalous dipole moments. We have also enriched
implementation of event weight for  $q \bar q \to \tau^-\tau^+ $ process with the possibility to study
impact of anomalous dipole moments on transverse spin correlations.

{There is important issue about the QCD and EW corrections; 
  those corrections have been extensively studied, for example,  in
  Refs.~\cite{Alwall:2014hca, Frederix:2018nkq, Degrande:2011ua, Darme:2023jdn}, in which progress 
	in automated calculation of matrix elements has been achieved.  Of course, for precise 
	SM simulation of the $\tau$-pair production, these higher-order corrections need to be taken into account.
  In the present paper, these corrections are included in framework of improved Born approximation 
	\cite{Bardin:1999yd, Richter-Was:2018lld}}. 

{Also the effects of multi-parton interactions \cite{Gieseke:2012aik, Lincoln:2016fgq} and underlying
  events \cite{CMS:2015wcf, CDF:2001onq, ATLAS:2010kmf} may be of importance.
  Note that in general, evaluations of underlying events and multi-parton interaction effects are 
  investigated  in framework of experimental collaboration work. At the same time, experimental details of
  detection and background subtractions have to be taken into account. Let us stress again 
  that all these efforts are needed to assure robustness of the event sample.
  Only if its precision is established,  one can expect reweighting results to be meaningful.}

{The {\tt TauSpinner} reweighting algorithm for introduction of new effects  is expected to supplement events with
  rather small corrections: event weights close to 1.  That is why demand on precision of NP effects is not very strict,  
  provided that predictions   of the SM  are not compromised. 
  The use of the present paper algorithms may be considered as the first step of the work,
  as we assume that the impact of interactions resulting with dipole moments will not affect the SM part of the interaction sizably.
}


\vspace{0.2cm} 
\centerline{\it Outlook}

Reliability of reweighting at high energies requires some attention, but as we work for a discovery tool, 
not a precision tool, that may be
postponed to further studies. So far, tests when  photon radiation is removed with generation variable 
cuts were performed only.
That is suitable for Belle II applications, but for high energy applications, such as at the FCC, further 
work on selection cuts
to prevent degradation of the performance due to bremsstrahlung is needed. It will require careful 
future considerations for
detector acceptance and more focus on the details of the detector than on tool validation.
At the FCC, the bremsstrahlung photons lost in the beam pipe
may have $p_T$ comparable to the $\tau$-lepton mass. That means further studies
and program extensions are needed for such configurations. {For example, 
at the c.m. energy around the $Z$-boson peak presence of such photons is largely reduced.}  
However, the program can now be used for studies with no bremsstrahlung
events or events of rather soft photons with energy and/or $p_T$ sizably   
smaller than half of the $\tau$ mass. This is not a constraint for program use at Belle II energies, 
where bremsstrahlung photons lost in the beam pipe cannot have large $p_T$.
 At high energies, the so-called Mustraal frame \cite{Berends:1982ie,Richter-Was:2016avq,Richter-Was:2016mal} 
for weight calculation may  need to be used, as already proposed for evolution of future versions of {\tt TauSpinner}.

\vspace{0.4cm}

\centerline {\bf Acknowledgments}

\vspace{0.4cm}

A.Yu.K. acknowledges partial support by the National Academy of Sciences of Ukraine 
via the program  ``Participation in the international projects in high-energy and 
nuclear physics'' (project No. C-4/53-2023). {He is grateful to the Polish Academy of Sciences 
for financial support during his stay at the Institute of Nuclear Physics PAS.}  
This project was supported in part from funds of  the Polish National Science Centre
under Decisions  No.~DEC-2017/27/B/ST2/01391, No.~UMO-2022/01/3/ ST2/00027,  
and of COPIN-IN2P3 collaboration with Laboratoire d'Annecy de Physique des Particules (LAPP), Annecy.
This project was supported by the U.S. Department of Energy under research Grant No.~DE-SC0022350.

\bibliographystyle{unsrt}
\bibliography{g-g_tau-tau}

\end{document}